\RequirePackage{fix-cm}
\documentclass[aip,cha,reprint,superscriptaddress,floatfix]{revtex4-2}

\usepackage[T1]{fontenc}
\usepackage{mathptmx}
\usepackage{amsmath,amssymb,bm}
\usepackage{graphicx}
\usepackage{booktabs}
\usepackage{microtype}
\usepackage[dvipsnames]{xcolor}
\usepackage{hyperref}
\usepackage{mathtools}
\usepackage{centernot}
\usepackage{etoolbox}
\makeatletter
\patchcmd{\NAT@citexnum}{--\NAT@penalty}{-\NAT@penalty}{}%
  {\PackageError{manuscript}{Could not set citation range hyphen}{Check natbib range formatting.}}
\makeatother
\ExplSyntaxOn
\cs_set_protected:Npn \bibinfo #1#2
  {
    \str_if_eq:nnTF {#1} {pages}
      {
        \group_begin:
        \tl_set:Nn \l_tmpa_tl {#2}
        \tl_replace_all:Nnn \l_tmpa_tl {--} {-}
        \tl_use:N \l_tmpa_tl
        \group_end:
      }
      {#2}
  }
\ExplSyntaxOff
\hypersetup{
    colorlinks=true,
    citecolor=RoyalBlue,
    urlcolor=RoyalBlue,
    linkcolor=MidnightBlue,
    pdftitle={Beyond density snapshots: Dynamical memory in Chern-Simons reservoirs},
    pdfauthor={Jyotiranjan Beuria and Venkatesh H. Chembrolu}
}

\usepackage{tikz}
\usetikzlibrary{arrows.meta,positioning,calc,fit,backgrounds}

\definecolor{navy}{HTML}{17365D}
\definecolor{blue}{HTML}{2F6FB3}
\definecolor{bluefill}{HTML}{EEF5FC}
\definecolor{purple}{HTML}{7352A3}
\definecolor{purplefill}{HTML}{F4F0FA}
\definecolor{teal}{HTML}{2D8C8C}
\definecolor{tealfill}{HTML}{EEF8F7}
\definecolor{orange}{HTML}{D97A2B}
\definecolor{orangefill}{HTML}{FFF4E9}
\definecolor{yellow}{HTML}{D4A017}
\definecolor{yellowfill}{HTML}{FFF9E8}
\definecolor{muted}{HTML}{666666}

\tikzset{
arrow/.style={-{Latex[length=2mm]},line width=.6pt,draw=black!65},
bus/.style={line width=.6pt,draw=black!55},
section/.style={
  anchor=west,font=\sffamily\bfseries\small,text=white,
  rounded corners=1mm,inner xsep=2.2mm,inner ysep=1mm,
  minimum width=5cm,align=left},
stage/.style={
  rounded corners=1.5mm,line width=.65pt,
  minimum height=1.4cm,align=center,inner sep=2mm,
  font=\sffamily\scriptsize},
model/.style={
  rounded corners=1.5mm,line width=.65pt,
  text width=3.5cm,minimum height=2.35cm,
  align=left,inner sep=2.2mm,font=\sffamily\scriptsize},
pill/.style={
  rounded corners=2mm,draw=black!45,fill=black!5,
  inner xsep=3mm,inner ysep=1mm,
  font=\sffamily\bfseries\scriptsize},
grouphead/.style={
  rounded corners=1.2mm,text=white,align=center,
  inner ysep=1.2mm,font=\sffamily\bfseries\scriptsize},
protocol/.style={
  rounded corners=1.2mm,line width=.55pt,fill=white,
  text width=7.1cm,minimum height=1.15cm,
  align=left,inner sep=1.8mm,font=\sffamily\scriptsize},
outcome/.style={
  rounded corners=1.5mm,line width=.65pt,
  text width=7.1cm,minimum height=1.3cm,
  align=left,inner sep=2mm,font=\sffamily\scriptsize}
}

\newcommand{\pnum}[2]{%
\tikz[baseline=(X.base)]{
\node[circle,fill=#1,text=white,inner sep=1pt,
minimum size=3.5mm,font=\sffamily\bfseries\tiny] (X) {#2};}}

\newcommand{\cC}{\mathcal C}
\newcommand{\norm}[1]{\left\lVert #1\right\rVert}

\begin{document}

\title[Dynamical memory in Chern-Simons reservoirs]{Beyond density snapshots: Dynamical memory in Chern-Simons reservoirs}

\author{Jyotiranjan Beuria}
\affiliation{ManasAI Technology Private Limited, Delhi, India}
\author{Venkatesh H. Chembrolu}
\affiliation{Indian Institute of Technology Mandi,
Mandi, India}

\date{18 October 2026}

\begin{abstract}
How a reservoir encodes input history jointly in density and geometric degrees of freedom remains insufficiently understood. We address this mechanistic gap with a continuous density on an internal manifold coupled to an independently evolving Chern-Simons (CS) connection through a conserved current. On the same internal manifold, we compare reciprocal transport, instantaneous transverse reconstruction, local nonlinear feedback, and coupled CS evolution. These controls isolate how transport and independent connection evolution contribute to memory. We identify a current-generated exact connection component that the tested instantaneous reconstruction and scalar local closure miss. Resetting this component at fixed density and Gauss-law curl changes the future density response within the chosen gauge and drift law and the small effect persists under numerical refinement. We then distinguish dynamical memory from its readout accessibility. Although the four models have similar scalar-memory scores and density traces of pulse order, their circulating and longitudinal traces differ. With all measured features, coupled CS improves pulse-order decoding by \(6.5\pm1.9\) percentage points over the instantaneous-transverse control in paired seedwise comparisons, while remaining comparable to local feedback. Coupled CS shows less lag than the instantaneous and local-feedback controls on the first cycle, but more on subsequent cycles, revealing a dependence on preparation. We thus connect a field-theoretic mechanism of history retention to controlled tests of its dynamical role and computational use. The four-model framework provides a way to determine what an evolving transport field contributes when density snapshots and aggregate performance alone leave the memory mechanism unresolved.
\end{abstract}

\maketitle

\begin{quotation}
Reservoirs can achieve similar scores while retaining the past in different
ways. We compare four models to find out which internal variables carry
history and when that history becomes useful for computation. The models
leave similar traces in density, but differ in how circulation and
longitudinal motion record the order of inputs. We then change part of
the transport field while preserving density and observe a different
future response. Repeated driving also changes which model appears to
respond with less lag. Together, these tests show how controlled
comparisons can reveal memory mechanisms that a single performance
score leaves unresolved.
\end{quotation}

\section{Introduction}
A reservoir retains input history through its evolving state, but a good
prediction score does not tell us which internal variables carry that
history. Reservoir computing makes this state useful by training a
readout of otherwise fixed dynamics
\cite{Maass2002Liquid,JaegerHaas2004Science,Luko2009Reservoir}, drawing on
the broader role of activity-dependent states in temporal processing
\cite{Buonomano2009StateDependent}. Its applications include reproducing
chaotic attractors \cite{Pathak2017ChaoticAttractors}, predicting
spatiotemporal chaos \cite{Pathak2018ModelFree}, and separating
superimposed chaotic signals \cite{Krishnagopal2020SeparationChaos}.
We ask how such dependence on the past is encoded when the reservoir
contains both a density field and an independently evolving transport
field.

Existing theory gives us several ways to assess temporal processing.
Fading memory describes diminishing dependence on distant inputs
\cite{BoydChua1985FadingMemory}, and echo-state networks can approximate
discrete-time fading-memory filters with uniformly bounded inputs
\cite{Grigoryeva2018UniversalESN}.
Capacity measures quantify functions of past inputs recoverable by a linear readout
\cite{Dambre2012Capacity}, while generalized synchronization
\cite{Hart2024GeneralisedSynchronisations} and minimal models of emulation
failure \cite{Sato2024MinimalModelRC} clarify the relation between driven
states and their source dynamics. We build on these perspectives by
examining the field dynamics that retain history alongside the readouts
that recover it.

Physical reservoirs already exploit nonlinear materials and continuous
media \cite{Tanaka2019PhysicalRC}, with computational access shaped by
input and measurement interfaces \cite{Stepney2024PhysicalRCTutorial}.
Examples range from Ginzburg-Landau polariton lattices
\cite{Opala2019PolaritonReservoir} and liquid-film waves
\cite{Maksymov2023FilmReservoir} to neuronal response models
\cite{Panahi2025NeuromorphicRC} and bistable viscoelastic chains
\cite{Ghoshal2025ViscoelasticRC}. These studies establish that a reservoir
can have a spatially extended dynamical state. Our question concerns how
history is shared between coupled density and geometric degrees of
freedom within such a state.

Topological reservoirs bring this question into sharper focus. Skyrmion
proposals use nonlinear resistance \cite{Prychynenko2018MagneticSkyrmion}
and random magnetic textures \cite{Pinna2020SkyrmionReservoir}; experiments
use confined motion for Boolean operations \cite{Raab2022BrownianReservoir},
strain-controlled textures for waveform classification
\cite{Sun2023SkyrmionReservoir}, and time multiplexing for gesture
recognition \cite{Beneke2024GestureReservoir}. Magnetic textures and
ferroelectric domain walls provide several routes to nonlinear response
and memory \cite{EverschorSitte2024TopologicalReservoirs}. More directly,
Singh et al. assign slow leakage to harmonic edge modes in a
Hodge-projected reservoir and test their role through cycle and projection
ablations \cite{Singh2026HodgeProjected}. Identifying the state responsible
for memory is therefore essential to understanding what a reservoir's
geometric structure contributes.

Here we develop a mechanistic, field-theoretic account of memory in a
continuous density fluctuation \(\rho(c,t)\) coupled to a gauge connection
\(a(c,t)\). We prescribe an internal manifold \(\cC\), whose coordinates
need not be physical positions, and use this input-conditioned field
state to describe internal context. Inputs redistribute the signed
density fluctuation, while the connection determines transport through
a specified drift law. Our earlier work introduced this construction to
study nonreciprocal collective dynamics, circulation, and hysteresis
\cite{Beuria2026TopologicalFlux}. We now investigate which aspects of
these dynamics store history and whether a readout can use them.

Chern-Simons (CS) coupling provides a useful separation between a density
constraint and connection evolution. Its foundations include
topologically massive gauge theory \cite{Deser1982TopologicallyMassive},
the topological character of pure CS theory \cite{Witten1989Jones}, and effective descriptions of
the fractional quantum Hall effect \cite{Zhang1989FQHE}. In our classical
model, Gauss law fixes the connection's curl from density, while the
spatial CS equations evolve the connection with the conserved current.
We make the driven density equation compatible with these equations by
adding a longitudinal current correction that incorporates injection and
relaxation without changing the density evolution. Nonlinear transport
then generates an exact, or gradient, connection component that the
instantaneous transverse reconstruction misses.

We test the contribution of this evolved state using four matched models:
reciprocal transport, instantaneous transverse reconstruction, local
nonlinear feedback, and coupled CS evolution. Matching matters because
prediction quality depends on reservoir realization
\cite{Haluszczynski2019GoodBadPredictions} and evaluation protocol
\cite{Wringe2025Benchmarks}. We also reset the exact connection component
at a checkpoint while preserving density and Gauss-law curl. The subsequent
density responses separate under identical inputs, demonstrating that
this component affects future dynamics within the chosen gauge and drift
law. This intervention establishes a dynamical role beyond the tested
density closures; it does not rule out every possible density-based
representation.

We assess computational usefulness separately through delayed-input
reconstruction and pulse-order decoding. All four models retain scalar
information and distinguish pulse order, while coupled CS has no resolved
scalar-memory advantage and does not outperform local feedback on
pulse-order decoding. We keep the domain topology fixed and the harmonic
sector zero. Our contribution is thus an account of gauge-organized
dynamical memory and its readout accessibility; we do not isolate a
computational benefit from topology or establish topological protection.

We define the field reservoir and derive its gauge-constrained dynamics
in Secs.~\ref{sec:context_memory} and \ref{sec:gauge_dynamics}. We introduce
the four controls in Sec.~\ref{sec:controlled_models} and describe the
numerics and diagnostic protocols in Secs.~\ref{sec:numerics},
\ref{sec:reservoir_protocols}, and \ref{sec:gauge_protocols}. We present
the results in Sec.~\ref{sec:reservoir_results} and conclude in
Sec.~\ref{sec:conclusion}.

\section{Field reservoir and local memory}
\label{sec:context_memory}

A physical reservoir has a state \(z(t)\), fixed internal dynamics,
temporal input \(u(t)\), measured features \(R[z(t)]\), and a trained
linear readout only:
\begin{equation}
 \dot z=F[z,u(t)],\qquad
 \widehat y=b+\bm w^{\mathsf T}R[z(t)].
 \label{eq:reservoir}
\end{equation}
In the present system \(z=(\rho,a)\), where \(\rho(c,t)\) is a
zero-mean density fluctuation and \(a(c,t)\) is a spatial gauge
one-form on the domain \(\cC\). We use \(c\) to label a position on this
domain, not an equivalence class of reservoir states. The reservoir state
is the full field configuration \(z\); \(\rho\) is a signed fluctuation,
not a probability density over contexts. We prescribe the domain geometry,
input profiles, and field dynamics, and train only the readout. We distinguish
two local notions of memory.  
\begin{itemize}
    \item Density memory is a
residual difference in \(\rho\) after different input histories.
\item Geometric local memory is a difference in circulation, divergence, or
other flow observables reconstructed from \(a\).  
\end{itemize}
We measure these state variables directly and use a linear readout to test
whether a simple observation protocol can recover information about past inputs.

This separation between fixed dynamics, measured features, and a trained
readout is essential to the reservoir-computing tests developed below.  The
input history must first be encoded by the autonomous response of the physical
state; the feature map then determines which parts of that response are
experimentally accessible, while the readout tests whether the encoded
information can be recovered without modifying the reservoir itself.  Keeping
these roles distinct also makes the comparison between models mechanistic: a
difference in performance can be traced to the available density or geometric
state variables rather than to model-specific training of the internal
dynamics.  The delayed-input and pulse-order protocols of
Sec.~\ref{sec:reservoir_protocols} therefore use the same dynamical states both
for direct memory diagnostics and for constructing the feature families given
to the linear readout. We summarize the study design and controlled model hierarchy in Fig.~\ref{fig:study_design}. 

\begin{figure*}[!t]
\centering
\resizebox{\textwidth}{!}{%
  \begin{tikzpicture}

\node[
  stage,
  draw=blue,
  fill=bluefill,
  text width=3.48cm,
  align=left,
  anchor=north west
] (input) {
  \textcolor{blue}{\textbf{Structured drive}}\\[.5mm]
  $u_\ell(t)\rho_{{\rm in},\ell}(\bm c)$
};

\node[
  stage,
  draw=purple,
  fill=purplefill,
  text width=3.48cm,
  align=left,
  right=3.2mm of input
] (state) {
  \textcolor{purple}{\textbf{Reservoir fields}}\\[.5mm]
  $\rho(\bm c,t)$ on periodic $\mathcal C$; model-dependent $\bm v,a$
};

\node[
  stage,
  draw=teal,
  fill=tealfill,
  text width=3.88cm,
  align=left,
  right=3.2mm of state
] (dynamics) {
  \textcolor{teal}{\textbf{Fixed dynamics}}\\[.5mm]
  $\partial_t\rho+\nabla\!\cdot\bm j=0$
};

\node[
  stage,
  draw=orange,
  fill=orangefill,
  text width=3.48cm,
  align=left,
  right=3.2mm of dynamics
] (features) {
  \textcolor{orange}{\textbf{Measured features}}\\[.5mm]
  $\rho,\bm v,\omega,\theta$
};

\draw[arrow] (input) -- (state);
\draw[arrow] (state) -- (dynamics);
\draw[arrow] (dynamics) -- (features);

\node[section] at ([yshift=-5.5mm]input.south west)
  {CONTROLLED FOUR-MODEL HIERARCHY};

\node[
  model,
  draw=blue,
  fill=bluefill,
  text width=3.55cm,
  align=left,
  below=9.5mm of input.south west,
  anchor=north west
] (m1) {
  \colorbox{blue}{\textcolor{white}{\textbf{\strut 1}}}\quad
  \textcolor{blue}{\textbf{Reciprocal gradient}}\\[1mm]
  $\bm v_{\rm rec}=-\gamma\nabla(-\nabla^2)^{-1}g\rho$\\[1mm]
  \textbf{Flow:} longitudinal only\\[.4mm]
  \textbf{State:} instantaneous $\rho(t)$
};

\node[
  model,
  draw=orange,
  fill=orangefill,
  text width=3.55cm,
  align=left,
  right=3.3mm of m1
] (m2) {
  \colorbox{orange}{\textcolor{white}{\textbf{\strut 2}}}\quad
  \textcolor{orange}{\textbf{Transverse control}}\\[1mm]
  $\bm v_0=-\gamma J\nabla(-\nabla^2)^{-1}g\rho$\\[1mm]
  \textbf{Flow:} circulating only\\[.4mm]
  \textbf{State:} $a_{\rm c}[\rho(t)]$
};

\node[
  model,
  draw=teal,
  fill=tealfill,
  text width=3.55cm,
  align=left,
  right=3.3mm of m2
] (m3) {
  \colorbox{teal}{\textcolor{white}{\textbf{\strut 3}}}\quad
  \textcolor{teal}{\textbf{Local feedback}}\\[1mm]
  $\bm v_{\rm loc}=(1+\eta\rho)\bm v_0$\\[1mm]
  \textbf{Flow:} circulating + longitudinal\\[.4mm]
  \textbf{State:} instantaneous $\rho(t)$
};

\node[
  model,
  draw=purple,
  fill=purplefill,
  line width=.85pt,
  text width=3.55cm,
  align=left,
  right=3.3mm of m3
] (m4) {
  \colorbox{purple}{\textcolor{white}{\textbf{\strut 4}}}\quad
  \textcolor{purple}{\textbf{Fully coupled CS}}\\[1mm]
  $\bm v=-\gamma a,\quad\partial_ta=gJ\bm j$\\[1mm]
  \textbf{Flow:} circulating + longitudinal\\[.4mm]
  \textbf{State:} independently evolved $a(t)$
};

\coordinate (foundation) at ($(input.south)!0.5!(features.south)$);
\coordinate (modelbus) at ($(foundation)+(0,-3.4mm)$);

\draw[bus] (foundation) -- (modelbus);
\draw[arrow] (modelbus) -| (m1.north);
\draw[arrow] (modelbus) -| (m2.north);
\draw[arrow] (modelbus) -| (m3.north);
\draw[arrow] (modelbus) -| (m4.north);

\node[
  font=\fontsize{6.8}{7.8}\selectfont,
  text=muted,
  fill=white,
  inner sep=1pt
] at (modelbus) {
  same $\mathcal C$, drive, numerics, initial states and seeds
};

\coordinate (modelsouth) at ($(m2.south)!0.5!(m3.south)$);

\node[pill,below=6mm of modelsouth] (analysisfork)
  {ANALYSIS FRAMEWORK};

\draw[arrow] (modelsouth) -- (analysisfork.north);

\node[
  grouphead,
  fill=blue,
  text width=7.55cm,
  align=center
] (rh) at ($(analysisfork)+(-4.20cm,-1.30cm)$) {
  RESERVOIR-COMPUTING PROTOCOLS
};

\node[
  grouphead,
  fill=purple,
  text width=7.55cm,
  align=center
] (gh) at ($(analysisfork)+(4.20cm,-1.30cm)$) {
  GAUGE-DYNAMICS PROTOCOLS
};

\coordinate (split) at ($(analysisfork.south)+(0,-3mm)$);
\draw[bus] (analysisfork.south) -- (split);
\draw[arrow] (split) -| (rh.north);
\draw[arrow] (split) -| (gh.north);

\node[
  protocol,
  draw=blue!55,
  text width=7.35cm,
  align=left,
  below=2.2mm of rh
] (r1) {
  \pnum{blue}{1}\quad
  \textbf{Scalar fading memory}\hfill $MC(d),\ C$\\[-.2mm]
  \textcolor{muted}{Recover past random inputs and test decay with delay.}
};

\node[
  protocol,
  draw=blue!55,
  text width=7.35cm,
  align=left,
  below=2.2mm of r1
] (r2) {
  \pnum{blue}{2}\quad
  \textbf{Matched pulse-order histories}\hfill
  $D_\rho,D_v,D_\omega,D_\parallel$\\[-.2mm]
  \textcolor{muted}{Measure endpoint separation, persistence, and held-out order decoding.}
};

\node[
  protocol,
  draw=blue!55,
  text width=7.35cm,
  align=left,
  below=2.2mm of r2
] (r3) {
  \pnum{blue}{3}\quad
  \textbf{Cyclic-drive density lag}\hfill $A_{\rm cyc}$\\[-.2mm]
  \textcolor{muted}{Measure rate-dependent failure of the return path to retrace the outward path.}
};

\node[
  protocol,
  draw=purple!55,
  text width=7.35cm,
  align=left,
  below=2.2mm of gh
] (g1) {
  \pnum{purple}{1}\quad
  \textbf{CS-equation residuals}\hfill $R_G,\ R_S$\\[-.2mm]
  \textcolor{muted}{Test Gauss-law consistency, spatial CS evolution, refinement, and noise robustness.}
};

\node[
  protocol,
  draw=purple!55,
  text width=7.35cm,
  align=left,
  below=2.2mm of g1
] (g2) {
  \pnum{purple}{2}\quad
  \textbf{Reconstruction and reset}\hfill $E_a,\ E_v$\\[-.2mm]
  \textcolor{muted}{Test transverse and local closures; reset the exact component at fixed density.}
};

\node[
  protocol,
  draw=purple!55,
  text width=7.35cm,
  align=left,
  below=2.2mm of g2
] (g3) {
  \pnum{purple}{3}\quad
  \textbf{Feedback-dependent cross-response}\hfill $\Delta R_\chi$\\[-.2mm]
  \textcolor{muted}{Measure the additional distant response generated by advective feedback.}
};

\begin{scope}[on background layer]
  \node[
    fit=(rh)(r1)(r3),
    draw=blue!45,
    fill=bluefill!45,
    rounded corners=1.6mm,
    inner sep=2.5mm,
    line width=.55pt
  ] {};

  \node[
    fit=(gh)(g1)(g3),
    draw=purple!45,
    fill=purplefill!45,
    rounded corners=1.6mm,
    inner sep=2.5mm,
    line width=.55pt
  ] {};
\end{scope}

\node[section] at ([yshift=-5.6mm]r3.south west)
  {INTERPRETATION};

\node[
  outcome,
  draw=yellow,
  fill=yellowfill,
  text width=7.7cm,
  align=left,
  below=9.8mm of r3
] (o1) {
  \textbf{Shared reservoir behavior}\\[.5mm]
  All four show fading memory, order-sensitive density, and finite-time lag.
};

\node[
  outcome,
  draw=purple,
  fill=purplefill,
  line width=.85pt,
  text width=7.7cm,
  align=left
] (o2) at (o1 -| g3) {
  \textbf{What coupled CS adds}\\[.5mm]
  Resetting the evolved exact component changes future density response.
};

\draw[arrow] (r3.south) -- (o1.north);
\draw[arrow] (g3.south) -- (o2.north);

\end{tikzpicture}
}
\caption{Study design and controlled model hierarchy. We use the same
spatial domain, input protocol, numerical scheme, initial
conditions, and seeds in all four models, and vary how density couples to
the velocity or gauge connection. We test scalar memory, pulse-order
dependence, and cyclic-drive lag with the reservoir-computing protocols.
With the gauge-dynamics protocols, we test Chern-Simons consistency,
snapshot reconstruction, connection reset, and feedback-dependent
cross-response. We then separate properties shared by the field
reservoir family from effects that require an independently evolving
gauge connection.}
\label{fig:study_design}
\end{figure*}

\section{Gauge-constrained field dynamics}
\label{sec:gauge_dynamics}

\subsection{The Chern-Simons formalism}
\label{subsec:cs}
Let \(j^\mu=(\rho,j^1,j^2)\) be a conserved current,
\begin{equation}
 \partial_\mu j^\mu
 =
 \partial_t\rho+\nabla_i j^i=0.
 \label{eq:continuity}
\end{equation}
We introduce an Abelian gauge field \(a_\mu\) governed by the
Chern-Simons action
\begin{equation}
 S[a;j]
 =
 \frac{\kappa}{4\pi}
 \int_{\mathcal{M}_3} d^3x\,
 \epsilon^{\mu\nu\lambda}
 a_\mu\partial_\nu a_\lambda
 -
 \int_{\mathcal{M}_3} d^3x\,
 a_\mu j^\mu ,
 \label{eq:cs_action}
\end{equation}
where \(\epsilon^{012}=+1\). Here, $\mathcal{M}_3=\mathbb{R}_t\times\mathcal{C}$ denotes the three-dimensional manifold formed by time and the two-dimensional spatial domain $\mathcal{C}$. Accordingly, $d^3x=dt d^2x$ with $t\in\mathbb{R}_t$ and $x \in \mathcal{C}$.
 Varying the action with respect to \(a_\mu\), we obtain
\begin{equation}
 \frac{\kappa}{2\pi}
 \epsilon^{\mu\nu\lambda}
 \partial_\nu a_\lambda
 =
 j^\mu,
 \qquad
 g\equiv\frac{2\pi}{\kappa}.
 \label{eq:cs}
\end{equation}

We use the Chern-Simons term for its standard role as a first-order,
metric-independent gauge response in two spatial dimensions \cite{Deser1982MassivePRL,Deser1982TopologicallyMassive,
Witten1989Jones,Zhang1989FQHE}.
In temporal gauge \(a_0=0\), we write Eq.~\eqref{eq:cs} in terms of the coupling $g$ as
\begin{align}
 \partial_1a_2-\partial_2a_1&=g\rho,                    \label{eq:gauss}\\
 \partial_ta_1&=g j^2,\qquad
 \partial_ta_2=-g j^1.                                 \label{eq:spatial}
\end{align}
In the overdamped limit, we neglect inertia and assume that the
drift responds instantaneously and linearly to the gauge one-form \(a\).
We therefore write
\begin{equation}
 \bm v=-\gamma a,\qquad \gamma>0.
 \label{eq:drift}
\end{equation}
Here \(\gamma\) is the mobility. Temporal gauge leaves transformations
\(a\mapsto a+\nabla\chi(c)\) with time-independent \(\chi\).
Equation~\eqref{eq:drift} changes under these transformations: it is a
constitutive law for a chosen representative, not a gauge-invariant
coupling. Temporal gauge, the initial representative, and the zero-harmonic
restriction therefore form part of this phenomenological reservoir model.

We express the spatial Chern-Simons equations
\begin{equation}
 \partial_t a_1=g j^2,
 \qquad
 \partial_t a_2=-g j^1
\end{equation}
in the compact form
\begin{equation}
 \partial_t a=gJ\bm j,
 \qquad
 J=
 \begin{pmatrix}
 0&1\\
 -1&0
 \end{pmatrix}.
\end{equation}
We retain these field equations from our previous work~\cite{Beuria2026TopologicalFlux}.

\subsection{Modelling external drive}

Inputs \(u_\ell(t)\) couple to prescribed profiles
\(\rho_{{\rm in},\ell}(c)\), deforming the internal density before we measure its transport and relaxation
through the readout.

We begin with a driven density equation in which the phenomenological
transport current is
\begin{equation}
 \bm j_{\rm p}
 =
 \alpha\rho\bm v-D\nabla\rho .
 \label{eq:precurrent}
\end{equation}
Here, \(\alpha\) controls the strength of advective transport generated by
the velocity \(\bm v\), while \(D>0\) is the diffusion coefficient.
The parameter \(\alpha\) is purely phenomenological and is unrelated to the
Chern-Simons coupling.

To model external driving, we supplement the transport dynamics with a
phenomenological source term consisting of linear relaxation and prescribed
input injection. We therefore choose
\begin{equation}
 s(c,t)
 =
 -\tau\rho(c,t)
 +
 \varepsilon\sum_{\ell=1}^{2}
 u_\ell(t)\rho_{{\rm in},\ell}(c),
 \label{eq:source}
\end{equation}
where \(\tau>0\) is the relaxation rate, \(\varepsilon\) sets the input
strength, \(u_\ell(t)\) are externally prescribed input signals, and
\(\rho_{{\rm in},\ell}(c)\) are their spatial profiles on the spatial
domain \(\cC\). 
The term \(-\tau\rho\) relaxes the density toward the reference
state \(\rho=0\), while \(u_\ell(t)\rho_{{\rm in},\ell}(c)\) injects the
$\ell$th input signal with a prescribed spatial profile. We choose this source phenomenologically; we do not derive it from the
Chern-Simons action.

The phenomenological transport current \(\bm j_{\rm p}\) and source
\(s\) define the driven density equation
\begin{equation}
 \partial_t\rho+\nabla\cdot\bm j_{\rm p}=s.
 \label{eq:driven_density}
\end{equation}
This equation is not itself a continuity equation because the source
\(s\) locally injects or removes density. However, the
Chern-Simons field equation requires the full spacetime current
\(j^\mu=(\rho,\bm j)\) to satisfy
\begin{equation}
 \partial_t\rho+\nabla\cdot\bm j=0.
 \label{eq:current_conservation_requirement}
\end{equation}

To represent the driven dynamics using such a conserved current, we
introduce a longitudinal current correction \(\nabla\psi\), where
\(\psi\) solves
\begin{equation}
 \nabla^2\psi=-s,
 \label{eq:poisson_completion}
\end{equation}
and define
\begin{equation}
 \bm j=\bm j_{\rm p}+\nabla\psi.
 \label{eq:completion}
\end{equation}
It then follows that
\begin{equation}
 \nabla\cdot\bm j
 =
 \nabla\cdot\bm j_{\rm p}+\nabla^2\psi
 =
 \nabla\cdot\bm j_{\rm p}-s.
\end{equation}
Substituting into Eq.~\eqref{eq:driven_density}, we recover
\begin{equation}
 \partial_t\rho+\nabla\cdot\bm j=0,
 \label{eq:completed_continuity}
\end{equation}
so the completed current reproduces exactly the original driven density
evolution while remaining compatible with the Chern-Simons constraint.

On a closed spatial domain \(\cC\), the Poisson equation
\(\nabla^2\psi=-s\) is solvable only when \(s\) has zero spatial mean.
Indeed,
\begin{equation}
 \int_{\cC}\nabla^2\psi\,dV
 =
 \int_{\partial\cC}\nabla\psi\cdot\bm n\,dS
 =
 0,
\end{equation}
because \(\partial\cC=\varnothing\), i.e., a closed manifold has no boundary surface. Therefore, a necessary solvability
condition is
\begin{equation}
 \int_{\cC}s\,dV=0.
 \label{eq:source_zero_mean}
\end{equation}
For the source specified in Eq.~\eqref{eq:source}, this condition holds
when the initial density and each input profile have zero spatial mean.
To check this, we integrate Eq.~\eqref{eq:driven_density} over \(\cC\) and obtain
\begin{equation}
 \frac{d}{dt}\int_{\cC}\rho\,dV
 =
 -\tau\int_{\cC}\rho\,dV
 +
 \varepsilon\sum_{\ell=1}^{2}u_\ell(t)
 \int_{\cC}\rho_{{\rm in},\ell}\,dV.
\end{equation}
Here we used the divergence theorem, $ \int_{\cC}\nabla\cdot\bm j_{\rm p}\,dV = \int_{\partial\cC}\bm j_{\rm p}\cdot\bm n\,dS = 0,$ since the closed manifold \(\cC\) has no boundary.
Hence, if
\begin{equation}
 \int_{\cC}\rho(c,0)\,dV=0,
 \qquad
 \int_{\cC}\rho_{{\rm in},\ell}(c)\,dV=0,
\end{equation}
then \(\int_{\cC}\rho\,dV=0\) and consequently
\(\int_{\cC}s\,dV=0\) for all times. Physically, the source may rearrange
positive and negative density deviations within \(\cC\), but it does not
change their net spatial integral.

The solution of Eq.~\eqref{eq:poisson_completion} is determined only up
to an additive constant, since \(\psi\) and \(\psi+C\) generate the same
current correction. We remove this ambiguity by imposing
\begin{equation}
 \int_{\cC}\psi\,dV=0.
 \label{eq:psi_zero_mean}
\end{equation}
This normalization does not affect \(\nabla\psi\); it merely selects a
unique scalar potential.

We choose the gradient completion as a modelling prescription rather than
derive it from microscopic transport. Among all corrections
\(\bm q\) satisfying
\begin{equation}
 \nabla\cdot\bm q=-s,
\end{equation}
the longitudinal choice \(\bm q=\nabla\psi\) has minimum \(L^2\) norm.

Since \(\nabla^2\psi=-s\), it can be written as
\begin{equation}
 \bm q=\nabla\psi+\bm w,
 \qquad
 \nabla\cdot\bm w=0.
\end{equation}
$\nabla\psi$ is the longitudinal part, while $\bm w$ is the transverse, divergence-free part. The longitudinal component changes the local density through its divergence; the transverse component only redistributes or circulates current without changing $\nabla\cdot q$.
On the closed manifold \(\cC\), integration by parts gives
\begin{equation}
 \int_{\cC}\nabla\psi\cdot\bm w\,dV
 =
 -\int_{\cC}\psi\,\nabla\cdot\bm w\,dV
 =0.
\end{equation}
Thus the longitudinal and divergence-free components are orthogonal, and
\begin{equation}
 \|\bm q\|_{L^2}^{2}
 =
 \|\nabla\psi\|_{L^2}^{2}
 +
 \|\bm w\|_{L^2}^{2}
 \geq
 \|\nabla\psi\|_{L^2}^{2}.
\end{equation}
Hence \(\nabla\psi\) is the minimum-\(L^2\)-norm correction compatible with
the prescribed divergence.

Adding a divergence-free current would change the CS evolution. We fix
this freedom by choosing the minimum-norm completion.

\subsection{Density-determined and evolving connection components}
\label{subsec:exact_memory}

On the periodic square with zero harmonic component, Gauss law implies
\begin{equation}
 \begin{aligned}
 a&=a_{\rm c}[\rho]+b,\qquad b=\nabla\chi(c,t),\\
 a_{\rm c}[\rho]&=gJ\nabla(-\nabla^2)^{-1}\rho.
 \end{aligned}
 \label{eq:connection_split}
\end{equation}
The transverse part is fixed by density. We project the spatial CS law
onto the exact sector using
\(P_\parallel=\nabla(\nabla^2)^{-1}\nabla\cdot\) and obtain
\begin{equation}
 \partial_t b=gP_\parallel J\bm j
 =-\alpha\gamma gP_\parallel J(\rho a).
 \label{eq:exact_memory}
\end{equation}
The diffusion and source-completion currents are gradients; their rotations
have zero longitudinal projection. Thus \(b\), initially zero, accumulates
through the nonlinear advective current and vanishes throughout when
\(\alpha=0\). This identifies a history integral beyond the tested
transverse reconstruction, without excluding every possible functional
of density. Removing \(b\) preserves Gauss law but changes the constitutive
velocity and potentially the subsequent density response. Such a reset is a
physical intervention within the fixed representative, not a gauge-invariant
memory measurement.

\section{Controlled field models}
\label{sec:controlled_models}

We design the controls to separate the dynamical ingredients of the CS
reservoir. All four models occupy the same periodic domain, and we keep
the harmonic sector zero throughout. We vary the constitutive coupling
and the availability of an independent connection, rather than changing
the topology of the domain. The reciprocal model supplies a
potential-driven baseline. With the instantaneous-transverse model, we
retain the density-sourced curl and circulating drift but remove the
independently evolving connection. With local feedback, we allow the
density to modulate that drift nonlinearly, testing a simple alternative
to retaining the connection's history. Comparing these controls with
coupled CS lets us separate transverse transport, local feedback, and
current-driven connection memory.

The distinction matters when interpreting a positive memory score.
If the reciprocal model retains a past input, that result cannot by
itself identify a CS mechanism. If the instantaneous or local control
matches a coupled-CS readout, the task does not resolve a benefit from
the independent connection. We therefore supplement the model comparison
with a reset of its exact component at fixed density and curl. This
intervention tests a component of the state directly, while leaving the
domain topology and harmonic sector unchanged. We do not interpret the
four models as distinct topological phases.

We use the same source term, diffusion, relaxation, input protocol, grid,
initial condition, and phenomenological transport current in all models:
\begin{equation}
 \bm j_{\rm p}
 =
 \alpha\rho\bm v-D\nabla\rho.
\end{equation}
We also use the same conserved completion,
\begin{equation}
 \bm j=\bm j_{\rm p}+\nabla\psi,
 \qquad
 \nabla^2\psi=-s,
\end{equation}
so that
\begin{equation}
 \partial_t\rho+\nabla\cdot\bm j=0.
\end{equation}
The comparison therefore changes only the constitutive relation between
the density, the gauge field, and the drift velocity. This provides a
controlled hierarchy in which reciprocal response, transverse response,
local nonlinear feedback, and independent gauge-field memory can be
introduced one at a time.

\paragraph{Reciprocal-gradient control.}
We first define a reciprocal control in which the density generates
an ordinary scalar potential,
\begin{equation}
 -\nabla^2\phi=g\rho,
 \label{eq:control_poisson}
\end{equation}
and choose the induced drift to follow the potential gradient:
\begin{equation}
 \bm v_{\rm rec}
 =
 -\gamma\nabla\phi
 =
 -\gamma g\,\nabla(-\nabla^2)^{-1}\rho.
 \label{eq:reciprocal}
\end{equation}
Because the inverse Laplacian is nonlocal, the velocity at a given point
depends on the density distribution over the full spatial domain. The
flow is nevertheless purely gradient-directed, and hence
\begin{equation}
 \nabla\times\bm v_{\rm rec}
 =
 -\gamma\,\nabla\times\nabla\phi
 =
 0.
\end{equation}
This model therefore retains nonlocal density-mediated transport while
excluding transverse circulation. We use it as the baseline for identifying the effect of the
Chern-Simons-induced rotation of the drift.

\paragraph{Instantaneous transverse control.}
We use the scalar potential \(\phi\) determined by
Eq.~\eqref{eq:control_poisson}.
Its gradient,
\begin{equation}
 \nabla\phi=(\partial_1\phi,\partial_2\phi),
\end{equation}
points in the direction of the steepest increase of \(\phi\). To obtain a
transverse response, we rotate this gradient by \(90^\circ\) using
\begin{equation}
 J=
 \begin{pmatrix}
  0&1\\
  -1&0
 \end{pmatrix}.
\end{equation}
Thus,
\begin{equation}
 a_{\rm c}
 =
 J\nabla\phi
 =
 (\partial_2\phi,-\partial_1\phi).
 \label{eq:ac_reconstruction}
\end{equation}
We define the corresponding drift by
\begin{equation}
 \bm v_0=-\gamma a_{\rm c}.
 \label{eq:leading}
\end{equation}

The rotated field is perpendicular to the original gradient because
\begin{equation}
 \nabla\phi\cdot a_{\rm c}
 =
 (\partial_1\phi)(\partial_2\phi)
 -
 (\partial_2\phi)(\partial_1\phi)
 =
 0.
\end{equation}
Hence \(a_{\rm c}\) points along the level curves of \(\phi\), producing
circulation around density-induced structures rather than motion directly
along the potential gradient.

The field is also divergence-free:
\begin{equation}
 \nabla\cdot a_{\rm c}
 =
 \partial_1\partial_2\phi
 -
 \partial_2\partial_1\phi
 =
 0,
\end{equation}
where we use equality of mixed derivatives. Its curl, however, is
nonzero:
\begin{align}
 \partial_1a_{{\rm c},2}
 -
 \partial_2a_{{\rm c},1}
 &=
 \partial_1(-\partial_1\phi)
 -
 \partial_2(\partial_2\phi)
 \nonumber\\
 &=
 -\nabla^2\phi
 =
 g\rho.
\end{align}
Therefore, \(a_{\rm c}\) satisfies the Chern-Simons Gauss-law (Eq. \ref{eq:gauss}) constraint
at every instant. Since it is reconstructed directly from the current
density \(\rho(t)\), this control retains the transverse spatial response
but does not possess an independently evolving gauge-field memory. Solving Eq.~\eqref{eq:control_poisson} gives
\begin{equation}
 \phi(t)=g(-\nabla^2)^{-1}\rho(t),
\end{equation}
and hence
\begin{equation}
 a_{\rm c}(t)
 =
 J\nabla\phi(t)
 =
 gJ\nabla(-\nabla^2)^{-1}\rho(t)
\end{equation}
and \begin{equation}
 \bm v_0(t)
 =
 -\gamma gJ\nabla(-\nabla^2)^{-1}\rho(t).
\end{equation}
We thus reconstruct \(a_{\rm c}\) entirely from the instantaneous density:
we first invert the Poisson equation, then rotate the resulting gradient
by \(90^\circ\).

\paragraph{Local nonlinear-feedback control.}
To isolate the effect of state-dependent feedback from that of independent
gauge-field memory, we introduce the phenomenological closure
\begin{equation}
 \bm v_{\rm loc}
 =
 (1+\eta\rho)\bm v_0.
 \label{eq:localclosure}
\end{equation}
Here, \(\eta\) is the feedback-control parameter. It determines how strongly
the instantaneous density modifies the amplitude of the transverse
drift: positive \(\eta\) enhances the local response where \(\rho>0\) and
suppresses it where \(\rho<0\), while \(\eta=0\) recovers the instantaneous
transverse control,
\begin{equation}
 \bm v_{\rm loc}\big|_{\eta=0}=\bm v_0.
\end{equation}
Equivalently,
\begin{equation}
 \bm v_{\rm loc}
 =
 \bm v_0+\eta\rho\,\bm v_0,
\end{equation}
so the second term is a local nonlinear feedback correction whose strength
can be varied continuously through \(\eta\).

This control is important because nonlinear feedback and dynamical memory
can produce similar history-dependent responses. Equation~\eqref{eq:localclosure}
adds nonlinear state dependence while keeping the field instantaneous:
\(\bm v_{\rm loc}(t)\) is determined entirely by \(\rho(t)\) and
\(\bm v_0(t)\), with no independently evolving gauge variable. Comparing
this model with the fully coupled Chern-Simons model therefore tests
whether the observed effects arise from local nonlinear amplification alone
or require memory stored in the gauge-field dynamics. 

\paragraph{Comparison with the fully coupled model.}
We compare these three controls with the fully coupled Chern-Simons model introduced in Sec.~\ref{subsec:cs}. Unlike the controls, the full model
evolves the gauge field independently through the conserved current and
can develop the exact component in Eq.~\eqref{eq:connection_split}.
The comparison tests what the specified instantaneous controls omit,
separating reciprocal transport, transverse circulation, local nonlinear
feedback, and evolving connection memory.

The four models differ in one constitutive ingredient at a time:
\begin{align}
 \bm v_{\rm rec}
 &=
 -\gamma\nabla(-\nabla^2)^{-1}g\rho,
 \\
 \bm v_0
 &=
 -\gamma J\nabla(-\nabla^2)^{-1}g\rho,
 \\
 \bm v_{\rm loc}
 &=
 (1+\eta\rho)\bm v_0,
 \\
 \bm v
 &=
 -\gamma a,
 \qquad
 \partial_ta=gJ\bm j.
 \label{eq:coupled_velocity_summary}
\end{align}
The reciprocal control tests nonlocal density-mediated dynamics without
circulation. The instantaneous transverse control adds circulation but
removes independent gauge memory. The local closure adds nonlinear
density feedback while remaining instantaneous. The coupled model alone
retains transverse response, nonlinear current feedback, and a gauge field
that integrates the history of the conserved current.

\section{Numerical implementation}
\label{sec:numerics}

We evolve all four models using the same spatial discretization,
time-integration scheme, external drive, initial-condition ensemble, and
random seeds. This lets us attribute differences in the measured responses
to the constitutive dynamics. Here we describe the common implementation;
we introduce the task-specific protocols in
Secs.~\ref{sec:reservoir_protocols} and \ref{sec:gauge_protocols}.

\subsection{Domain and spectral discretization}

We take the spatial domain to be the flat periodic torus
\(\cC=[0,L)^2\), with side length \(L=2\pi\) and coordinates
\(\bm c=(c_1,c_2)\). The dynamical
variables are the zero-mean density fluctuation
\(\rho(c_1,c_2,t)\), the drift field \(\bm v\), the conserved current
\(\bm j\), and, in the coupled model, the spatial gauge connection
\(a=(a_1,a_2)\). We denote spatial averages by \(\langle\cdot\rangle\) and define the root-mean-square norm as
\[
\norm{f}_2=\left\langle |f|^2\right\rangle^{1/2}.
\]

Unless stated otherwise, we run the simulations on a \(16\times16\)
uniform grid. We compute spatial derivatives and Poisson inversions
spectrally using the integer Fourier wave vectors \(\bm k=(n_1,n_2)\).
For the instantaneous controls, we define the scalar potential through
the zero-mean Poisson problem
\begin{equation}
 -\nabla^2\phi=g\rho,
 \qquad \langle\phi\rangle=0,
 \qquad
 \widehat\phi_{\bm k}=\frac{g\widehat\rho_{\bm k}}{|\bm k|^2}
 \quad(\bm k\neq0).
 \label{eq:numerical_poisson}
\end{equation}
The reciprocal and instantaneous-transverse velocities are then
\(\bm v_{\rm rec}=-\gamma\nabla\phi\) and
\(\bm v_0=-\gamma J\nabla\phi\), respectively. The rotated field has
curl \(g\rho\) and hence satisfies the Gauss-law constraint at that
instant. For the coupled-CS model, we evolve the connection from the conserved
current and use Eq.~\eqref{eq:numerical_poisson} only during initialization
and diagnostic reconstruction.

We evaluate nonlinear products, in particular the advective term \(\rho\bm v\), pseudospectrally. Such products generate
wave numbers larger than those resolved by the discrete grid, which can
be spuriously folded onto lower Fourier modes. We suppress this aliasing
with the standard componentwise two-thirds truncation. Since the Nyquist
wave number of an \(N\)-point grid is \(N/2\), retaining two thirds of
the resolved bandwidth gives the cutoff
\((2/3)(N/2)=N/3\). After each nonlinear multiplication we therefore set
to zero all Fourier coefficients satisfying
\[
|n_1|>N/3
\qquad\text{or}\qquad
|n_2|>N/3.
\]
For the \(16\times16\) grid, the integer modes retained by this criterion
are \(|n_1|,|n_2|\leq5\). Applying the cutoff separately to both
components is the tensor-product form of the two-thirds rule appropriate
to the square Fourier grid.

For scalar Poisson problems, we set the \(\bm k=0\) Fourier coefficient
of the inverse Laplacian to zero, consistent with our zero-mean convention.
In the coupled gauge-field evolution, we separately remove the spatially
uniform component of the vector connection, since the Gauss-law constraint
on the torus does not fix it.

\subsection{Time integration and parameters}

We integrate in time using a fourth-order Runge-Kutta scheme with time step
\[
\Delta t=0.015.
\]
We sample the external inputs and hold them fixed during each integration
step. In the principal simulations, we use
\begin{align*}
 g&=0.45, &
 \gamma&=0.65, &
 D&=0.045,\\
 \tau&=0.12, &
 \varepsilon&=0.20, &
 \alpha&=0.70.
\end{align*}
Here \(g\) is the density-gauge coupling, \(\gamma\) converts the
connection into an overdamped drift, \(D\) is the diffusivity, \(\tau\)
is the linear density-relaxation rate, \(\varepsilon\) sets the input
amplitude, and \(\alpha\) controls the strength of advective
current feedback. For the local nonlinear control we use
\(\eta=0.35\), except in the closure-fitting analysis, where we determine \(\eta\) explicitly from the simulated data.

\subsection{Inputs, initial conditions, and ensembles}

We introduce two input channels through the fixed zero-mean profiles
\begin{align}
 \rho_{{\rm in},1}
 &\propto
 \cos c_1+\cos(c_1+c_2),
 \nonumber\\
 \rho_{{\rm in},2}
 &\propto
 \sin c_2+\sin(c_1-c_2),
 \label{eq:inputs}
\end{align}
each normalized to unit root-mean-square amplitude. Equal weights avoid a
fitted shape parameter. These are the only source profiles; protocols vary
only \(u_1(t)\) and \(u_2(t)\). Later Gaussian masks and packets are readouts
and probes, not alternative inputs.

When we need a nonzero random initial density, we construct it as a sum
of eight cosine modes with \(1\leq n_1\leq3\),
\(-3\leq n_2\leq3\), and independently sampled coefficients and phases.
We subtract its spatial mean and normalize it to the root-mean-square
amplitude stated for the corresponding diagnostic. All four models use
the same realization for a given seed. We initialize the coupled connection as
\(a(0)=gJ\nabla(-\nabla^2)^{-1}\rho(0)\), using
Eq.~\eqref{eq:numerical_poisson}, while each control is initialized from
the same density field.

Unless stated otherwise, we compute ensemble averages over ten
predetermined seeds paired across the four models. Shaded bands show the standard error of the
mean (SEM), \(\sigma_{\rm sample}/\sqrt{10}\), where
\(\sigma_{\rm sample}\) is the sample standard deviation across seeds.
We define the controlled mechanism hierarchy in Sec.~\ref{sec:controlled_models}; the
shared realizations used here isolate those constitutive differences.

\section{Reservoir-computing protocols}
\label{sec:reservoir_protocols}

We test three aspects of reservoir behavior: recovery of past scalar
inputs, discrimination of input order, and rate-dependent lag. The
pulse-order analysis includes endpoint separation and held-out linear
classification. Throughout,
\(u_\ell(t)\) in Eq.~\eqref{eq:source} is the scalar amplitude of the fixed
spatial profile \(\rho_{{\rm in},\ell}(\bm c)\); it is not a component of
the evolving density \(\rho(\bm c,t)\).

\subsection{Scalar fading memory}
We first test whether the reservoir retains information about a past scalar
input. We use only the first input channel, so \(u_2(t)=0\) and
\(u_1(t)=u(t)\). The source term becomes
\begin{equation}
 s(\bm c,t)=-\tau\rho(\bm c,t)
 +\varepsilon u(t)\rho_{{\rm in},1}(\bm c).
 \label{eq:scalar_memory_source}
\end{equation}

The input consists of independent random amplitudes
\(u_n\sim\mathcal U[-1,1]\). We hold each amplitude constant for one symbol interval,
\begin{equation}
 u(t)=u_n,\qquad
 nT_s\leq t<(n+1)T_s,\qquad T_s=3\Delta t.
 \label{eq:symbol_hold}
\end{equation}
The reservoir is therefore driven by a sequence
\(\ldots,u_{n-2},u_{n-1},u_n\). If its dynamics has fading memory, its state
after receiving \(u_n\) should still contain some information about the
earlier input \(u_{n-d}\), with this information generally becoming weaker
as the delay \(d\) increases.

At the end of symbol interval \(n\), we summarize the density field by
\begin{equation}
 \begin{aligned}
 \bm x_n&=\left(
 \bigl(\operatorname{Re}\widehat\rho_{\bm k},
       \operatorname{Im}\widehat\rho_{\bm k}\bigr)_{\bm k\in\mathcal K},
 \langle\rho^2\rangle\right)
 \in\mathbb R^9,\\
 \mathcal K&=\{(1,0),(0,1),(1,1),(1,-1)\}.
 \end{aligned}
 \label{eq:scalar_memory_features}
\end{equation}
The set \(\mathcal K\) is not unique: in principle, one could use any number
of Fourier modes scattered across wave-vector space. We deliberately retain
only the lowest nonzero axial and diagonal modes because they capture robust,
large-scale density variations while keeping the readout low-dimensional.
For a real density field, the corresponding negative modes contain no
independent information. The moment \(\langle\rho^2\rangle\) adds the overall density variation to the feature set. This restricted feature set
tests whether memory is accessible from simple, preselected observables rather than from an exhaustive representation of the field. We standardize the features using means and standard deviations calculated from the training data, producing \(\bm z_n\); we then apply the same transformation to the test data.

For each delay \(d=1,\ldots,16\), we train a separate linear readout to
reconstruct the input that occurred \(d\) symbol intervals before the current
reservoir state. The readout is
\begin{align}
 \widehat u_{n,d}&=b_d+\bm\beta_d^{\mathsf T}\bm z_n,
 \label{eq:memory_readout}\\
 (b_d,\bm\beta_d)&=\mathop{\arg\min}_{b,\bm\beta}
 \mathcal L_d(b,\bm\beta),\nonumber\\
 \mathcal L_d(b,\bm\beta)
 &=\sum_{n\in\mathcal T_d}
 \left[u_{n-d}-b-\bm\beta^{\mathsf T}\bm z_n\right]^2
 \nonumber\\
 &\quad+10^{-4}\norm{\bm\beta}_2^2,
 \label{eq:memory_ridge}
\end{align}
where \(\mathcal T_d\) denotes the valid training indices for delay \(d\).
At each time \(n\), \(\bm z_n\) is the present reservoir feature vector,
\(u_{n-d}\) is the input presented \(d\) symbols earlier, and
\(\widehat u_{n,d}\) is its reconstruction from the present reservoir state
alone. For example, for \(d=3\), we train the readout using pairs of the form
\[
(\bm z_3,u_0),\quad
(\bm z_4,u_1),\quad
(\bm z_5,u_2),\quad \ldots .
\]
We fit a different pair of coefficients \((b_d,\bm\beta_d)\) for every delay \(d\). The squared-error term trains the readout to reproduce the past
input, while the ridge term \(10^{-4}\norm{\bm\beta}_2^2\) stabilizes the fit
when reservoir features are correlated. We do not penalize the intercept \(b_d\). Since \(u_n\) is treated as a continuous-valued target, this is a
regression task rather than a classification task.

After fitting the readout for a given delay \(d\), we keep its coefficients
fixed and evaluate it on a completely independent test sequence. We quantify
memory performance by
\begin{equation}
 MC(d)=\operatorname{corr}^2[u_{n-d},\widehat u_{n,d}],\qquad
 C=\sum_{d=1}^{16}MC(d),
 \label{eq:memorycapacity}
\end{equation}
where we compute the Pearson correlation across all valid test times \(n\) for that fixed delay. Thus, \(MC(d)\) is a single number obtained by comparing
two vectors over the test sequence: the true delayed inputs \(u_{n-d}\) and
their reconstructions \(\widehat u_{n,d}\). If the present reservoir state
retains information about the input from \(d\) symbols earlier, these two
vectors vary together and \(MC(d)\) approaches one. If that past input is no
longer linearly recoverable from the present state, the correlation is small
and \(MC(d)\) approaches zero. The total memory score \(C\) sums the
linearly accessible memory over delays \(d=1,\ldots,16\). One delay unit
corresponds to three RK4 integration steps.

We train and test on separate input streams. We begin each run from zero
density and apply 50 washout symbols to move the reservoir away from its
initial condition. We exclude these symbols from fitting and evaluation,
then reindex the retained portion of each stream from \(n=0\). The training run contains 350 retained
symbols, whereas the independent test run contains 250 retained symbols.

For a delay \(d\), the first \(d\) retained reservoir states cannot be used
because their targets \(u_{n-d}\) would refer to symbols lying inside the
discarded washout interval. The usable training pairs are therefore
\[
(\bm z_d,u_0),\quad
(\bm z_{d+1},u_1),\quad
\ldots,\quad
(\bm z_{349},u_{349-d}),
\]
giving \(350-d\) training examples. We treat the independent test run in exactly the same way, giving
\[
(\bm z_d,u_0),\quad
(\bm z_{d+1},u_1),\quad
\ldots,\quad
(\bm z_{249},u_{249-d}),
\]
or \(250-d\) test examples. We keep the coefficients fitted on the training sequence fixed and compute
Eq.~\eqref{eq:memorycapacity} only from these \(250-d\) independent test pairs. All four models
receive exactly the same training and test streams, allowing their memory
curves to be compared directly.

\subsection{Matched pulse-order histories}
We next test whether the final reservoir state depends on the order of two
input events. We apply the same two pulses in opposite sequences. Because the initial state, total input, and final input are
matched, a difference between the two final states means that the reservoir
retains the path taken rather than only the accumulated drive. This provides
a direct test of order-sensitive memory; the density, circulation, and
longitudinal diagnostics below identify which dynamical channel carries that
memory.

Both source channels in Eq.~\eqref{eq:source} are active, and
\(\bm u=(u_1,u_2)\) denotes their amplitudes. A pulse on channel 1 means
\(\bm u=(A_{\rm p},0)\), so that the injected part of the source is
\(\varepsilon A_{\rm p}\rho_{{\rm in},1}\); a pulse on channel 2 similarly
uses \(\bm u=(0,A_{\rm p})\) and injects
\(\varepsilon A_{\rm p}\rho_{{\rm in},2}\).

Let \(W\), \(G\), and \(R\) denote the pulse width, intervening gap, and
final relaxation time, all measured in RK4 steps. For compactness,
\([\bm q]_m\) denotes a constant input vector \(\bm q\) held for \(m\)
steps, and \(\mathbin{\Vert}\) denotes temporal concatenation: the block to
its right begins when the block to its left ends. It is not a product. The
two protocols are
\begin{align}
 \bm u_A&=[(A_{\rm p},0)]_W\mathbin{\Vert}[(0,0)]_G
           \mathbin{\Vert}[(0,A_{\rm p})]_W\mathbin{\Vert}[(0,0)]_R,\nonumber\\
 \bm u_B&=[(0,A_{\rm p})]_W\mathbin{\Vert}[(0,0)]_G
           \mathbin{\Vert}[(A_{\rm p},0)]_W\mathbin{\Vert}[(0,0)]_R.
 \label{eq:pulse_protocols}
\end{align}
Thus protocol A applies channel 1 before channel 2, whereas protocol B
reverses their order. We observe the reservoir after the final zero-input relaxation interval, at
\(t_f=(2W+G+R)\Delta t\). At this time,
\begin{align}
 \bm u_A(t_f)&=\bm u_B(t_f)=0,\nonumber\\
 \int_0^{t_f}\bm u_A\,dt
 &=\int_0^{t_f}\bm u_B\,dt
 =A_{\rm p}W\Delta t\,(1,1),
 \label{eq:balanced}
\end{align}
where the integrals are componentwise. The two histories therefore have
the same duration, final input, and total input delivered to each channel;
only the order differs. Any difference between their final states is
therefore a history-dependent response rather than a consequence of unequal
input exposure.

We use \(W\in\{8,12,16\}\),
\(A_{\rm p}\in\{0.50,0.75,1.00,1.25\}\), \(G=5\), and \(R=12\). For a
time step \(\Delta t=0.015\), the three widths correspond to pulse durations
\(0.12\), \(0.18\), and \(0.24\). They probe short, intermediate, and longer
exposure, while the four amplitudes probe the response across increasing
input strength. This compact factorial grid tests whether an order effect
persists across both duration and strength rather than at one selected drive;
the values are protocol choices, not optimized model parameters. For a given
seed, both protocols and all four models start from the same random density
field of root-mean-square amplitude \(0.04\).

\subsection{Endpoint separation of matched histories}

The two protocols in Eq.~\eqref{eq:pulse_protocols} have the same total
duration, the same final input, and the same time-integrated input in each
channel; they differ only in the order of the two pulses. Comparing their
final states therefore tests whether the reservoir retains information about
input order. This complements scalar memory: scalar recall measures how long
a past input amplitude remains recoverable, whereas path separation asks
whether two histories with the same overall input content but different order
lead to distinguishable reservoir states. Resolving the separation into
density, circulating, and longitudinal components further indicates which
physical channel carries the history dependence.

To probe local geometric structure, we define two Gaussian observation masks
\(w_1(\bm x)\) and \(w_2(\bm x)\) on the square domain of side \(L\). A mask
is a fixed spatial weighting function: for any field \(f(\bm x)\),
\(\langle w_i f\rangle\) is a weighted spatial average that emphasizes the
region around the center of \(w_i\). The masks therefore act only as local
observation windows and do not affect the dynamics or the applied inputs.

Both masks have standard deviation \(0.15L\) and unit spatial mean,
\[
 w_i(\bm x)=
 \frac{
 \exp[-|\bm x-\bm c_{w,i}|^2/(2\sigma_w^2)]
 }{
 \left\langle
 \exp[-|\bm x-\bm c_{w,i}|^2/(2\sigma_w^2)]
 \right\rangle
 },
 \qquad
 \sigma_w=0.15L,
\]
with centers
\[
 \bm c_{w,1}=(L/4,L/3),
 \qquad
 \bm c_{w,2}=(3L/4,3L/5).
\]
We fix these locations in advance, choosing well-separated points that
avoid simple square symmetries. The endpoint diagnostics in this subsection use only
\(w=w_1\) as a fixed local probe, whereas the reservoir feature vector
introduced later contains observables extracted with both \(w_1\) and
\(w_2\).

At the final time \(t_f\), we compare the states reached under the two
pulse-order protocols using
\begin{align}
 D_\rho&=\norm{\rho_A-\rho_B}_{2},\nonumber\\
 D_v&=\norm{\bm v_A-\bm v_B}_{2},                      \label{eq:distances}\\
 D_\omega&=\left|\langle
 w(\nabla\times\bm v_A-\nabla\times\bm v_B)\rangle\right|,
 \nonumber\\
 D_\parallel&=\left|\langle
 w(\nabla\cdot\bm v_A-\nabla\cdot\bm v_B)\rangle\right|.
 \nonumber
\end{align}
Here, subscripts \(A\) and \(B\) denote the two pulse-order protocols.
\(D_\rho\) and \(D_v\) are global \(L^2\) distances between the final density
and velocity fields. In contrast, \(D_\omega\) and \(D_\parallel\) are local
quantities evaluated with \(w_1\): they measure differences in the
circulating response, through \(\nabla\times\bm v\), and in the longitudinal
(divergence) response, through \(\nabla\cdot\bm v\), respectively. Since these quantities have different physical units, we do not compare
their absolute magnitudes directly.

We compute all four raw quantities for every model. To make small
model-to-model differences easier to resolve, we additionally form paired,
dimensionless contrasts relative to the coupled-CS model. For density and
circulation,
\begin{equation}
 \Delta_X^{(m)}=100\,
 \frac{D_X^{(m)}-D_X^{({\rm CS})}}
 {D_X^{({\rm CS})}+10^{-20}},\qquad X\in\{\rho,\omega\},
 \label{eq:pairedcontrast}
\end{equation}
while for the longitudinal response we use
\begin{equation}
 Q_\parallel^{(m)}=
 \frac{D_\parallel^{(m)}}{D_\parallel^{({\rm CS})}+10^{-20}}.
 \label{eq:longratio}
\end{equation}
The superscript \({\rm CS}\) denotes the coupled-CS value for the same seed
and pulse pair, and \(10^{-20}\) regularizes an extremely small denominator.
Thus, \(\Delta_X^{(m)}=0\) and \(Q_\parallel^{(m)}=1\) denote equality with
coupled CS. Because the factor \(100\) is included in
Eq.~\eqref{eq:pairedcontrast}, \(\Delta_X\) is already percentage valued; no
additional factor of \(100\) is applied when it is reported. For the amplitude table, we first average the raw distances over \(W\)
within each seed at fixed \(A_{\rm p}\). For the persistence table, we
average them over \(W\) and \(A_{\rm p}\) within each seed at fixed time.
We then form the paired quantities and compute their mean and SEM across seeds. We do not introduce a
separate normalized contrast for
\(D_v\), because \(D_v\) combines circulating and longitudinal velocity
differences that are resolved more directly by \(D_\omega\) and
\(D_\parallel\).

To test whether the order-dependent separation persists after the drive is
removed, we continue each matched-history pair after the second pulse with
\(u_1=u_2=0\), while the relaxation term \(-\tau\rho\) in
Eq.~\eqref{eq:source} remains active. Over the following 48 RK4 steps
(\(0.72\) time units), we define the post-drive time \(t_{\rm rel}\), with
\(t_{\rm rel}=0\) immediately after the second pulse, and record \(D_\rho\),
\(D_\omega\), and
\(D_\parallel\), together with the paired quantities in
Eqs.~\eqref{eq:pairedcontrast} and \eqref{eq:longratio}. A nonzero value
during this interval indicates that the two pulse orders remain
distinguishable for a finite time after the drive has ended; it does not
imply permanent storage. The standard endpoint at \(t_f\), which includes the
\(R=12\) final-relaxation steps in Eq.~\eqref{eq:pulse_protocols}, corresponds
to \(t_{\rm rel}=R\Delta t=0.18\).

\subsection{Linear decodability of pulse order}

The endpoint distances above establish whether the two input histories
produce different states, but they do not show whether pulse order can be
recovered by a simple reservoir readout. We therefore test whether the order
label is linearly decodable from five feature families. Let
\(\mathcal K=\{(1,0),(0,1),(1,1),(1,-1)\}\) and define the eight-component
low-mode vector
\begin{equation}
 \mathcal F_{\mathcal K}[\phi]
 =\bigl(\operatorname{Re}\widehat\phi_{\bm k},
        \operatorname{Im}\widehat\phi_{\bm k}\bigr)_{\bm k\in\mathcal K},
 \label{eq:lowmodefeatures}
\end{equation}
with the modes ordered as listed in \(\mathcal K\). Writing
\(\omega=\nabla\times\bm v\) and \(\theta=\nabla\cdot\bm v\), the five
feature vectors are
\begin{align}
 \bm x_\rho&=\bigl(\mathcal F_{\mathcal K}[\rho],
                    \langle\rho^2\rangle\bigr)\in\mathbb R^9,\nonumber\\
 \bm x_\omega&=\bigl(\mathcal F_{\mathcal K}[\omega],
                    \langle\omega^2\rangle,
                    \langle w_1\omega\rangle,\langle w_2\omega\rangle\bigr)
                    \in\mathbb R^{11},\nonumber\\
 \bm x_\parallel&=\bigl(\mathcal F_{\mathcal K}[\theta],
                    \langle\theta^2\rangle,
                    \langle w_1\theta\rangle,\langle w_2\theta\rangle\bigr)
                    \in\mathbb R^{11},\nonumber\\
 \bm x_{\rm geom}&=(\bm x_\omega,\bm x_\parallel)\in\mathbb R^{22},
 \qquad
 \bm x_{\rm all}=(\bm x_\rho,\bm x_{\rm geom})\in\mathbb R^{31}.
 \label{eq:featurefamilies}
\end{align}
The families separate density, circulation, and longitudinal flow.
Low modes retain amplitude and phase, quadratic moments measure total
variation, and masked averages retain localized signed information.

For each run, we sample the selected feature vector \(\bm x_j\) at \(t_f\) and assign the label \(y_j=+1\) for protocol A or \(-1\) for protocol B.
On standardized training data, we fit
\begin{equation}
 f(\bm x)=b+\bm a^{\mathsf T}\bm x,
 \qquad
 \min_{b,\bm a}\sum_j[y_j-f(\bm x_j)]^2+\lambda\norm{\bm a}_2^2,
 \label{eq:pulse_classifier}
\end{equation}
and predict the pulse order from the sign of \(f\). Although the classifier
is fitted with a ridge squared-error loss, the target has only the two order
labels. Its held-out accuracy is
\begin{equation}
 \operatorname{Acc}=\frac{1}{N_{\rm test}}
 \sum_{j\in\mathrm{test}}
 \mathbf 1\!\left[\operatorname{sign}f(\bm x_j)=y_j\right].
 \label{eq:pulse_accuracy}
\end{equation}
Chance accuracy is \(1/2\). Each of 50 initial-condition seeds contributes
24 endpoints: two pulse orders for each of three widths and four amplitudes.
The 1200 observations therefore form 50 seed blocks.

We evaluate generalization by nested leave-one-seed-out validation. In each
outer fold, we hold out all 24 observations from one seed for testing and
use the remaining 49 seeds to provide
\(49\times24=1176\) training observations. Keeping complete seed blocks in a
single fold prevents information leakage between paired runs generated from
the same initial-condition realization. We estimate feature means and standard deviations only from the outer
training set and apply them unchanged to the held-out seed.

Within each outer training set, we select the ridge coefficient
\(\lambda\in\{10^{-6},10^{-5},\ldots,10^2\}\) by an inner
leave-one-seed-out validation over the 49 training seeds. We do not penalize
the intercept \(b\). After selecting \(\lambda\), we refit the classifier
on all 49 outer-training seeds and evaluate it on the single unseen seed.
Repeating this procedure over all 50 outer folds yields one held-out accuracy
for each seed.

We summarize held-out seed scores by their means and SEMs and pair model
differences by seed. The 1200 endpoints are not independent replicates, and
outer folds share training data, so these SEMs are descriptive ensemble
uncertainties rather than independent-fold significance tests.

\subsection{Cyclic-drive density lag}

To complement the pulse-order test discussed above, we examine how the density responds to a
cyclic input. If the outward and return branches differ at the same input
amplitude, then the present state retains information about the path taken.
We refer to this finite-time path dependence as cyclic-drive lag.

In Eq.~\eqref{eq:source}, we sweep the input amplitude \(u_1\) from
\(-1\) to \(1\) and back while keeping \(u_2=0.50\), so that the second
channel remains active throughout. The sweep is deterministic. Each branch
contains \(N_b\) uniformly spaced input values, with one value applied per
RK4 step of duration \(\Delta t=0.015\). The return branch visits the same
values in reverse order, so a complete cycle contains \(2N_b\) RK4 steps.

Since \(u_1\) changes by 2 over one branch, the nominal sweep rate is
\[
 r_s=\frac{2}{N_b\Delta t}.
\]
We use
\(N_b=160,\allowbreak 320,\allowbreak 640,\allowbreak 1280,\allowbreak 2560\), corresponding to
\(r_s=0.8333,\allowbreak 0.4167,\allowbreak 0.2083,\allowbreak 0.1042,\allowbreak 0.0521\), respectively. Successive
values therefore reduce the driving rate by a factor of two. This scan tests
whether the lag decreases under slower driving or whether slow sweeps instead
lead to accumulated lag or numerical instability. These rates are fixed
protocol choices, not fitted parameters.

We monitor
\[
 E_\rho=\langle\rho^2\rangle,
\]
which measures the overall strength of the density pattern and is defined for
all four models. We do not use vorticity here because it vanishes identically for the
reciprocal control. Denoting the outward and return branches by
\(E_\rho^+(u_1)\) and \(E_\rho^-(u_1)\), we quantify their separation by
\begin{equation}
 A_{\rm cyc}=\int_{-1}^{1}
 \left|E_\rho^+(u_1)-E_\rho^-(u_1)\right|\,du_1.
 \label{eq:hystarea}
\end{equation}
If the two branches coincide, \(A_{\rm cyc}=0\); larger values indicate
stronger finite-cycle lag.

Each run starts from a random density field with root-mean-square amplitude
\(0.08\), using the same set of seeds across models and sweep rates. There is
no pre-equilibration at \(u_1=-1\), so \(A_{\rm cyc}\) includes the effect of
initial relaxation. We therefore interpret it as a finite-cycle lag measure, not as
quasistatic hysteresis, multistability, or a topological invariant.

We classify runs that produce a nonfinite density or connection field as
numerically unstable and compute means and SEMs only from finite runs.
For the representative comparison, we use the middle value of the five-rate
scan, \(r_s=0.2083\). This corresponds to \(N_b=640\), so one full cycle
contains 1280 RK4 steps.

\section{Gauge-dynamics protocols}
\label{sec:gauge_protocols}

The reservoir protocols above establish observable memory and readout
performance. Here we isolate what requires an independently evolved gauge
connection. Three diagnostics answer successive questions. First,
\(R_G\) and \(R_S\) test whether the numerical solution satisfies the two CS
field equations. Second, \(E_a\) and \(E_v\) ask whether the evolved
connection and velocity can be reconstructed from the instantaneous density.
Third, \(\Delta R_\chi\) measures the change in a local cross response caused
by advective feedback after subtracting the background response. All tests
use the numerical scheme of Sec.~\ref{sec:numerics}; within a test, models
share the same seed, initial density, and drive.

All deterministic two-channel evolutions below use the common, nonoptimized
drive \(u_1(t)=\sin t\), \(u_2(t)=\cos(2t)\). The integer harmonics provide
reproducible excitation without fitted frequency parameters. The scalar-memory
test discussed earlier is different by design: independent random symbols ensure that delayed
input reconstruction measures information retained by the reservoir rather
than the predictability of a periodic drive.

\subsection{CS-equation residuals: \texorpdfstring{\(R_G\)}{R-G} and \texorpdfstring{\(R_S\)}{R-S}}

We begin from ten random density fields with root-mean-square amplitude
\(0.12\) and evolve each for four time units ($\approx 267$ RK4 steps) under the common drive defined
above.

Along these trajectories, we monitor the connection
\(a=(a_1,a_2)\), density \(\rho\), and conserved current
\(\bm j=(j^1,j^2)\). Because the discretized dynamics need not satisfy the
Gauss-law and spatial CS equations exactly, we quantify their violation by
the normalized residuals
\begin{align}
 R_G&=\frac{\norm{\partial_1a_2-\partial_2a_1-g\rho}_2}
 {\norm{g\rho}_2+10^{-14}},                              \label{eq:gaussres}\\
 R_S&=\frac{\norm{\partial_ta-g(j^2,-j^1)}_2}
 {\norm{g\bm j}_2+10^{-14}}.                            \label{eq:spatialres}
\end{align}
Here \(R_G\) measures violation of Gauss law, while \(R_S\) measures
deviation from the spatial CS evolution equation. The constant \(10^{-14}\)
only regularizes a vanishing denominator.

The Gauss-law residual \(R_G\) is meaningful only for the coupled-CS model.
The instantaneous and local transverse controls reconstruct
\(a_{\rm c}[\rho]\) directly from the density and therefore satisfy Gauss law
by construction, while the reciprocal model has no CS connection. By
contrast, \(R_S\) distinguishes an independently evolved connection from one
reconstructed instantaneously from the density. For coupled CS, we estimate \(\partial_t a\) by a trapezoidal finite difference. For the
instantaneous control,
\(a_{\rm c}=gJ\nabla(-\nabla^2)^{-1}\rho\), and hence
\(\partial_ta_{\rm c}=gJ\nabla(-\nabla^2)^{-1}\partial_t\rho\); this is
evaluated with Eq.~\eqref{eq:numerical_poisson} using \(\partial_t\rho\) in
place of \(\rho\). Thus the instantaneous-control derivative is implied by
the density evolution rather than evolved as an independent degree of
freedom. Using identical initial \(\rho\), \(a\), and drive isolates this
difference.

We additionally test numerical robustness by repeating the residual analysis
on \(12^2\), \(16^2\), and \(24^2\) grids and with the time step
\(\Delta t\) reduced by a factor of two. These refinement runs use the same
initial field and nominal endpoint \(t\simeq2\), with rounded step counts
and drive samples at each integration step.

As a separate robustness check, we evolve three realizations with weak
Gaussian perturbations of amplitude \(2\times10^{-4}\sqrt{\Delta t}\)
added to the density at every time step. We remove the spatial mean of each
perturbation, apply the dealiasing filter, and reconstruct the corresponding
connection increment using Eq.~\eqref{eq:numerical_poisson}. This keeps the
added perturbation consistent with Gauss law. These tests assess whether
the observed values of \(R_G\) and \(R_S\) are stable under changes in
spatial resolution, time-step size, and weak constraint-compatible perturbation.

\subsection{Snapshot reconstruction: \texorpdfstring{\(E_a\)}{E-a} and \texorpdfstring{\(E_v\)}{E-v}}

Constraint satisfaction alone does not determine agreement with a specified
snapshot reconstruction. We quantify that agreement by writing \(\bm v_{\rm CS}\equiv\bm v=-\gamma a\) for the
full coupled velocity in Eq.~\eqref{eq:coupled_velocity_summary}:
\begin{align}
 E_a&=\frac{\norm{a-a_{\rm c}[\rho]}_2}{\norm{a}_2+10^{-14}},
 &a_{\rm c}[\rho]&=gJ\nabla(-\nabla^2)^{-1}\rho,
 \label{eq:connection_mismatch}\\
 E_v&=\frac{\norm{\bm v_{\rm candidate}-\bm v_{\rm CS}}_2}
 {\norm{\bm v_{\rm CS}}_2+10^{-14}}.                   \label{eq:closureerror}
\end{align}
Here \(E_a\) is nontrivial only for coupled CS: it vanishes for a connection
reconstructed from the same density, while the reciprocal model has no CS
connection. The velocity error \(E_v\) instead compares the coupled velocity
with the instantaneous transverse field \(\bm v_0\) and its fitted local
closure in Eq.~\eqref{eq:localclosure}.

For ten seeds at
\(\alpha\in\{0,0.15,0.30,0.50,0.70,0.90\}\), we evolve 267 steps
(\(t_*=4.005\)) under the common drive and fit one scalar \(\eta\) to
minimize \(\norm{(1+\eta\rho)\bm v_0-\bm v_{\rm CS}}_2\) at each
endpoint separately. Comparison with \(\eta=0\) tests this local
multiplier under favorable fitting conditions; it does not exclude
arbitrary nonlocal or nonlinear density-based reconstructions.

To test dynamical relevance, we clone each coupled state at \(t_*\),
retain exactly the same density, and replace only the clone's connection
by \(a_{\rm c}[\rho]\). Both copies then evolve with identical future
inputs for \(1.005\) time units. We measure
\(\delta_\rho=\norm{\rho-\rho_{\rm reset}}_2/\norm{\rho}_2\).
The reset removes \(b\) while preserving curl, Gauss law, and the harmonic
restriction. It is an intervention on an admissible state, not evidence
for two naturally generated histories with identical endpoint density.

We perform the intervention with seeds 101 to 110 and repeat it at \(24^2\),
\(\Delta t/2\), \(\alpha=0,0.50,0.90\), and
\(\tau=0.09,0.15\), varying one parameter at a time. Refinement holds
inputs for the original \(0.015\)-unit intervals and matches physical
endpoints. The same grid and time refinements also test three consecutive
cycles at \(N_b=640\) without reinitialization, and the 50-seed,
31-feature pulse-order classifier, for the two instantaneous controls
and coupled CS.

\subsection{Feedback-dependent cross-response: \texorpdfstring{\(\Delta R_\chi\)}{Delta R-chi}}

We test whether a small density perturbation applied at one site produces a
measurable velocity response at a distant site, and whether that response is
modified by advective feedback through the gauge field. The probe is needed to
create a controlled local perturbation whose effect can be compared against an
otherwise identical unperturbed trajectory. Through Gauss law, the density
probe induces a compatible change in the connection \(a\), and hence in
\(\bm v=-\gamma a\). During the subsequent evolution, the modified velocity
changes the advective current
\(\bm j_{\rm adv}=\alpha\rho\bm v\), which then feeds back into \(a\).
We therefore measure a finite-time cross-susceptibility, rather than a
transport rate or propagation speed.

For each seed, we first evolve a random density field of rms amplitude
\(0.12\) for four time units (\(\approx267\) RK4 steps) under the common
drive. This prepares the background state
\((\rho^{(0)},a^{(0)})\). We then apply a localized probe at one of two
well-separated sites,
\(\bm c_1=(L/4,L/3)\) and
\(\bm c_2=(3L/4,3L/5)\):
\begin{equation}
 \begin{aligned}
 G_i(\bm c)&=\exp\!\left[-\frac{d_{\rm per}(\bm c,\bm c_i)^2}
 {2\sigma_{\rm probe}^2}\right],\\
 \delta\rho_i(\bm c)&=A_{\rm probe}
 \bigl[G_i(\bm c)-\langle G_i\rangle\bigr].
 \end{aligned}
 \label{eq:gaussian_probe}
\end{equation}
Here \(d_{\rm per}\) is the shortest distance on the periodic square and
\(\sigma_{\rm probe}=L/8\). Subtracting \(\langle G_i\rangle\) gives
\(\langle\delta\rho_i\rangle=0\), preserving total mass and compatibility
with periodic Gauss law and the zero-mode condition in
Eq.~\eqref{eq:numerical_poisson}.

The probe amplitude \(A_{\rm probe}=2\times10^{-3}\) sets the perturbation
size and is not a model parameter. Halving or doubling it changes the
normalized response by less than \(0.0005\%\), showing that the measured
response is effectively linear over this range. Probe \(i\) defines
\[
\rho^{(i)}=\rho^{(0)}+\delta\rho_i,
\]
and we reconstruct the corresponding connection increment \(\delta a_i\) with Eq.~\eqref{eq:numerical_poisson}, so that
\(a^{(i)}=a^{(0)}+\delta a_i\) remains Gauss-law compatible.

After applying the probe, we evolve the unperturbed and perturbed
trajectories for an additional \(0.12\) time units, corresponding to eight
RK4 steps, while holding the drive at its terminal value. The initial four-time-unit
evolution prepares the background; this shorter post-probe interval allows
the induced velocity change to modify \(\bm j_{\rm adv}\) and feed back into
the connection before we measure the response. We repeat the test for
\[
 \alpha\in\{0,0.05,0.10,0.20,0.35,0.50,0.70\}.
\]
The case \(\alpha=0\) removes the advective current and provides the
no-feedback reference. The instantaneous control instead reconstructs its
velocity directly from each density snapshot and does not evolve an
independent connection.

After the post-probe evolution, let \(\bm v^{(0)}\) denote the background
velocity and \(\bm v^{(i)}\) the velocity following probe \(i\). The
opposite-site responses per unit probe amplitude are
\begin{equation}
 \begin{aligned}
  \bm\chi_{12}&=
  \frac{\bm v^{(2)}(\bm c_1)-\bm v^{(0)}(\bm c_1)}{A_{\rm probe}},\\
  \bm\chi_{21}&=
  \frac{\bm v^{(1)}(\bm c_2)-\bm v^{(0)}(\bm c_2)}{A_{\rm probe}}.
 \end{aligned}
 \label{eq:cross_response}
\end{equation}
Thus \(\bm\chi_{12}\) is the response at site 1 to a probe at site 2, while
\(\bm\chi_{21}\) is the reverse response. For a purely antisymmetric
transverse response, these vectors cancel in their sum. Defining
\(\bm\Sigma_\chi=\bm\chi_{12}+\bm\chi_{21}\), we report
\begin{align}
 R_\chi&=\frac{\norm{\bm\Sigma_\chi}}
 {\norm{\bm\chi_{12}}+\norm{\bm\chi_{21}}+10^{-14}},\nonumber\\
 \Delta R_\chi(\alpha)&=
 \frac{\norm{\bm\Sigma_\chi(\alpha)-\bm\Sigma_\chi(0)}}
 {\norm{\bm\chi_{12}}+\norm{\bm\chi_{21}}+10^{-14}}.
 \label{eq:responsemetrics}
\end{align}
The Euclidean norm is taken over velocity components, with each denominator
evaluated at the same \(\alpha\) as its numerator. \(R_\chi\) measures the
total failure of forward-reverse cancellation, whereas
\(\Delta R_\chi\) subtracts the seedwise \(\alpha=0\) reference and isolates
the additional cross-response generated by advective feedback.

\section{Results: connection memory and reservoir behavior}
\label{sec:reservoir_results}

We first isolate the evolving connection component and its effect on future
density response, then assess equation consistency and the memory behavior
shared across the four models.

\subsection{Snapshot reconstruction: \texorpdfstring{\(E_a\)}{E-a} and \texorpdfstring{\(E_v\)}{E-v}}

\begin{figure*}[!ht]
 \includegraphics[width=\textwidth]{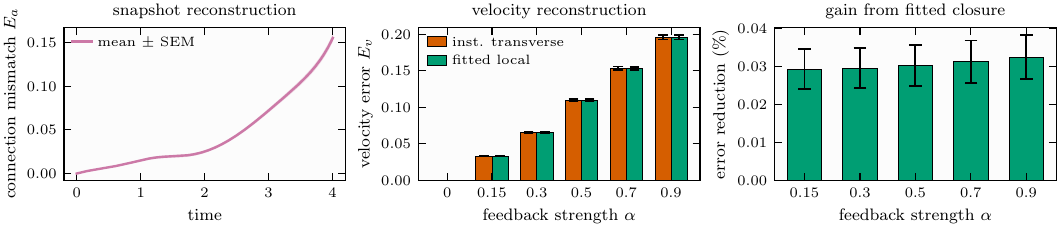}
 \caption{Reconstruction and closure at \(t_f\simeq4\). (a) Mean \(E_a(t)\)
 with SEM band at \(\alpha=0.70\). (b) Mean \(E_v\) for instantaneous
 transverse and fitted-local reconstruction, with SEM. (c) Their paired
 percentage difference (mean \(\pm\) SEM).}
 \label{fig:reconstruction}
\end{figure*}

We find a mean reconstruction mismatch \(E_a=0.155\) at \(t\simeq4\),
although both connections have the same curl to numerical precision
(Fig.~\ref{fig:reconstruction}(a)). The missing component is the exact field
\(b\) in Eq.~\eqref{eq:connection_split}, generated by the history of
the advective current in Eq.~\eqref{eq:exact_memory}.

We next test whether a fitted local multiplier can absorb this missing
component in the velocity reconstruction. At \(\alpha=0.90\), the instantaneous-transverse error
is \(0.19597\pm0.00301\), while the best local fit gives
\(0.19590\pm0.00301\) (Fig.~\ref{fig:reconstruction}(b)). The corresponding
reduction is only \(0.0325\%\pm0.0058\%\), despite a fitted coefficient of
\((-2.50\pm0.25)\times10^{-2}\) (Fig.~\ref{fig:reconstruction}(c)). The
closure in Eq.~\eqref{eq:localclosure} therefore scarcely improves this
snapshot fit; the result excludes this scalar local correction, not all
instantaneous density functionals.

We use the checkpoint intervention to test the consequence for future response.
Starting from identical densities at \(t_*=4.005\), removing \(b\) changes
the density after a further \(1.005\) units by
\(\delta_\rho=(1.77\pm0.17)\%\) (ten-seed mean \(\pm\) SEM).
The effect vanishes at \(\alpha=0\). Refinement to \(24^2\) changes the
mean \(E_a\) and \(\delta_\rho\) by less than \(0.02\%\) relative;
halving the integration step with fixed input holds leaves both unchanged
to the reported precision. When we vary \(\alpha\) from \(0.50\) to
\(0.90\) or \(\tau\) from \(0.09\) to \(0.15\) separately, we obtain
mean separations from \(0.92\%\) to \(2.85\%\), with
\(R_G<4.9\times10^{-13}\) throughout.
Thus the removed component affects density evolution within the chosen
constitutive model. This intervention does not establish reachability of
identical-density states from different natural input histories.

\begin{figure*}[!ht]
 \centering
 \includegraphics[width=0.8\textwidth]{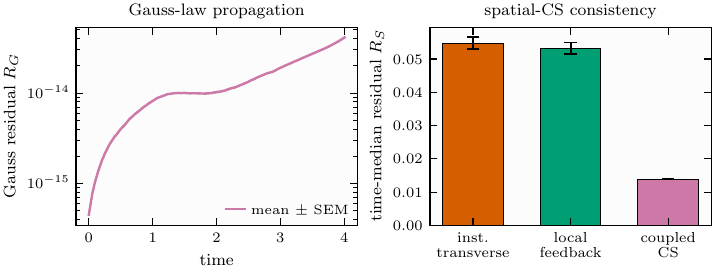}
 \caption{CS-equation residuals on the \(16^2\) grid with
 \(\Delta t=0.015\) and \(t_f\simeq4\). (a) Mean Gauss residual \(R_G\).
 (b) Mean seedwise time-median spatial residual \(R_S\).}
 \label{fig:residuals}
\end{figure*}

\subsection{CS-equation residuals: \texorpdfstring{\(R_G\)}{R-G} and \texorpdfstring{\(R_S\)}{R-S}}

Gauss law fixes the density-sourced curl, whereas the spatial CS equation
governs current-driven evolution. An instantaneous reconstruction can
satisfy the former without satisfying the latter.

We find that the coupled update propagates Gauss law to numerical
precision (Fig.~\ref{fig:residuals}(a)): the largest residual over ten seeds is
\(4.42\times10^{-14}\). In Fig.~\ref{fig:residuals}(b), the
ensemble-and-time median spatial residual is \(1.39\times10^{-2}\) for
coupled CS, compared with \(5.47\times10^{-2}\) for the
instantaneous-transverse reconstruction and \(5.32\times10^{-2}\) for local
feedback. Satisfying Gauss law at each density snapshot is therefore not
sufficient to reproduce the spatial CS evolution.

We report the refinement and perturbation checks for the residual protocol in Table~\ref{tab:checks}. Halving \(\Delta t\) reduces the
coupled spatial residual from \(1.07\times10^{-2}\) to
\(5.30\times10^{-3}\), indicating that its remaining value is dominated by
temporal discretization of the diagnostic. Grid refinement changes the final
density norm by only \(5.5\times10^{-6}\) between \(12^2\) and \(16^2\),
and by \(1.8\times10^{-6}\) between \(16^2\) and \(24^2\), relative to the
\(16^2\) result. Halving the time step changes the norm by \(0.77\%\),
but these original residual runs also change input sampling and the rounded
endpoint; the fixed-input tests above isolate integration refinement. Under
constraint-compatible noise, the largest relative change is
\(1.5\times10^{-4}\), and every run retains \(R_G<2.0\times10^{-14}\).
Thus the propagated constraint and density response are stable to the tested
grid, time-step, and weak-noise perturbations.

\begin{table*}[!ht]
\centering
\caption{Numerical validation of coupled CS. Unless indicated otherwise,
values use the \(16^2\) grid and \(\Delta t=0.015\). The two
spatial-residual convergence runs have nominal duration 2.}
\label{tab:checks}
\small
\begin{tabular}{@{}lc@{}}
\toprule
Diagnostic & Result\\
\midrule
Maximum \(R_G\), main run & \(4.42\times10^{-14}\)\\
Coupled spatial residual, \(\Delta t\) & \(1.07\times10^{-2}\)\\
Coupled spatial residual, \(\Delta t/2\) & \(5.30\times10^{-3}\)\\
\(\norm{\rho}_2\), \(N=12/16/24\) & \(0.261882/0.261881/0.261881\)\\
\(\norm{\rho}_2\), \(\Delta t/2\) & \(0.263887\)\\
\(\norm{\rho}_2\), three noise seeds & \(0.261844\) to \(0.261888\)\\
\bottomrule
\end{tabular}
\end{table*}

\begin{table*}[!ht]
\caption{Endpoint pulse-order contrasts versus amplitude, after \(R=12\)
zero-input steps. For each seed and \(A_{\rm p}\), we average raw distances
over \(W\in\{8,12,16\}\) before forming contrasts relative to coupled CS.
Entries are ten-seed means $\pm$ SEM; coupled CS is the exact reference.}
\label{tab:pathseparation}
\centering
\scriptsize
\begin{tabular}{@{}llcccc@{}}
\toprule
Observable & model & $A_{\rm p}=0.50$ & $A_{\rm p}=0.75$ & $A_{\rm p}=1.00$ & $A_{\rm p}=1.25$ \\
\midrule
density contrast $\Delta_\rho$ (\%) & reciprocal & $0.319\pm0.210$ & $0.369\pm0.206$ & $0.437\pm0.201$ & $0.523\pm0.197$ \\
 & inst. transverse & $-0.010\pm0.001$ & $-0.014\pm0.002$ & $-0.018\pm0.002$ & $-0.024\pm0.002$ \\
 & local feedback & $-0.011\pm0.002$ & $-0.014\pm0.002$ & $-0.018\pm0.002$ & $-0.024\pm0.003$ \\
 & coupled CS & $0$ (reference) & $0$ (reference) & $0$ (reference) & $0$ (reference) \\
\addlinespace
circulation contrast $\Delta_\omega$ (\%) & reciprocal & $-100.000$ ($D_\omega\simeq0$) & $-100.000$ ($D_\omega\simeq0$) & $-100.000$ ($D_\omega\simeq0$) & $-100.000$ ($D_\omega\simeq0$) \\
 & inst. transverse & $-0.009\pm0.002$ & $-0.011\pm0.002$ & $-0.014\pm0.002$ & $-0.017\pm0.002$ \\
 & local feedback & $0.132\pm0.114$ & $0.207\pm0.115$ & $0.281\pm0.115$ & $0.355\pm0.116$ \\
 & coupled CS & $0$ (reference) & $0$ (reference) & $0$ (reference) & $0$ (reference) \\
\bottomrule
\end{tabular}
\end{table*}

\begin{figure*}[!ht]
 \includegraphics[width=0.47\textwidth]{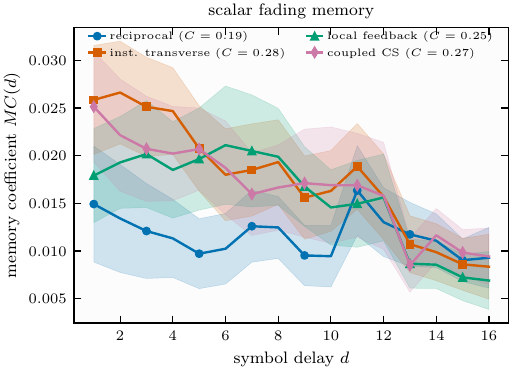}
 \hfill
 \includegraphics[width=0.47\textwidth]{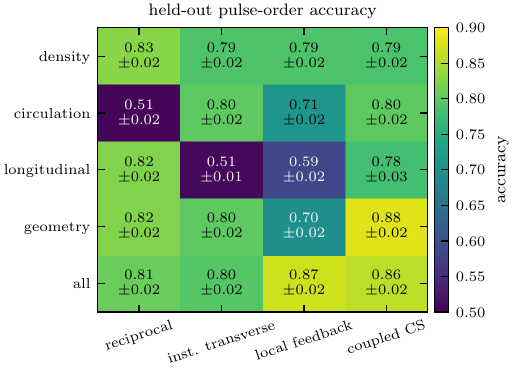}
 \caption{Held-out readout diagnostics. (a) Scalar-memory coefficient
 \(MC(d)\) for continuous delayed-input regression; curves and bands show the
 ten-seed mean and SEM, and legend values give \(C=\sum_dMC(d)\). (b) Binary
 pulse-order classification accuracy for the five feature families. Cells give
 the mean \(\pm\) SEM over 50 outer folds, with ridge penalties selected by
 nested seed-block validation; chance accuracy is \(0.5\).}
 \label{fig:readouts}
\end{figure*}

\subsection{Scalar fading memory}

We report the delayed-input reconstruction defined in Eq.~\eqref{eq:memorycapacity} in Fig.~\ref{fig:readouts}(a). For all four models, \(MC(d)\) is positive
at short delays and decreases overall across the 16-symbol window. The present
state therefore contains a linearly accessible but fading record of earlier
scalar inputs.

The corresponding capacities \(C\) are \(0.19\pm0.05\),
\(0.28\pm0.06\), \(0.25\pm0.06\), and \(0.27\pm0.06\) for reciprocal,
instantaneous-transverse, local-feedback, and coupled-CS dynamics,
respectively. The uncertainties overlap, so
we do not resolve a scalar-memory advantage for coupled CS. We recover
recent input amplitudes from all four evolving density fields.

\begin{table*}[tp]
\caption{Post-drive pulse-order contrasts. Here \(t_{\rm rel}=0\) is
immediately after the second pulse; the endpoint in
Table~\ref{tab:pathseparation} is at \(t_{\rm rel}=0.18\). For each seed and
time, we average raw distances over \(W\) and \(A_{\rm p}\) before
forming contrasts and ratios relative to coupled CS. Entries are ten-seed
means $\pm$ SEM; coupled CS is the exact reference.}
\label{tab:path}
\centering
\scriptsize
\begin{tabular}{@{}llcccc@{}}
\toprule
Observable & model & $t_{\rm rel}=0$ & $t_{\rm rel}=0.24$ & $t_{\rm rel}=0.48$ & $t_{\rm rel}=0.72$ \\
\midrule
density contrast $\Delta_\rho$ (\%) & reciprocal & $0.552\pm0.208$ & $0.404\pm0.200$ & $0.299\pm0.194$ & $0.222\pm0.189$ \\
 & inst. transverse & $-0.010\pm9.89\times10^{-4}$ & $-0.021\pm0.002$ & $-0.030\pm0.003$ & $-0.039\pm0.004$ \\
 & local feedback & $-0.010\pm0.002$ & $-0.021\pm0.003$ & $-0.030\pm0.003$ & $-0.039\pm0.004$ \\
 & coupled CS & $0$ (reference) & $0$ (reference) & $0$ (reference) & $0$ (reference) \\
\addlinespace
circulation contrast $\Delta_\omega$ (\%) & reciprocal & $-100.000$ ($D_\omega\simeq0$) & $-100.000$ ($D_\omega\simeq0$) & $-100.000$ ($D_\omega\simeq0$) & $-100.000$ ($D_\omega\simeq0$) \\
 & inst. transverse & $-0.008\pm0.001$ & $-0.016\pm0.002$ & $-0.023\pm0.003$ & $-0.029\pm0.004$ \\
 & local feedback & $0.285\pm0.119$ & $0.266\pm0.114$ & $0.249\pm0.110$ & $0.233\pm0.106$ \\
 & coupled CS & $0$ (reference) & $0$ (reference) & $0$ (reference) & $0$ (reference) \\
\addlinespace
longitudinal ratio $Q_\parallel$ & reciprocal & $159.646\pm61.489$ & $166.740\pm66.430$ & $171.572\pm69.205$ & $176.613\pm72.125$ \\
 & inst. transverse & numerical zero & numerical zero & numerical zero & numerical zero \\
 & local feedback & $0.755\pm0.329$ & $0.755\pm0.339$ & $0.742\pm0.337$ & $0.730\pm0.335$ \\
 & coupled CS & $1.000$ (reference) & $1.000$ (reference) & $1.000$ (reference) & $1.000$ (reference) \\
\bottomrule
\end{tabular}
\end{table*}

\subsection{Matched pulse-order histories}

We next follow the matched-history protocol of
Eq.~\eqref{eq:pulse_protocols}. Protocols A and B have the same duration,
final input, and channelwise integrated input in Eq.~\eqref{eq:balanced}; only
the pulse order changes. We first compare their endpoint states, then follow
the separation after the drive is removed, and finally ask whether a linear
readout can decode the order.

\subsubsection{Endpoint separation of matched histories}

We first evaluate the four raw distances in Eq.~\eqref{eq:distances} at the
standard endpoint \(t_f\), after the \(R=12\) zero-input steps. Unless stated
otherwise, a raw endpoint value quoted below is an unweighted mean over the
ten diagnostic seeds, all three pulse widths
\(W\in\{8,12,16\}\), and all four amplitudes
\(A_{\rm p}\in\{0.50,0.75,1.00,1.25\}\), with \(G=5\) and \(R=12\) fixed.
In the model order reciprocal, instantaneous transverse, local feedback, and
coupled CS, the grid-averaged density and velocity distances are
\begin{align}
 D_\rho&=(2.097,\,2.088,\,2.088,\,2.088)\times10^{-3},\nonumber\\
 D_v&=(5.103,\,5.093,\,5.104,\,5.102)\times10^{-4}.
 \label{eq:raw_endpoint_global}
\end{align}
Thus all four models retain a measurable endpoint trace of pulse order, while
the global density and velocity norms alone show little model discrimination.

Table~\ref{tab:pathseparation} resolves the endpoint comparison by pulse
amplitude. Each \(A_{\rm p}\) column holds the amplitude fixed, averages the
raw distances over the three widths within each seed, forms the paired
contrast, and then averages over the ten seeds. The reciprocal entries are
\(\Delta_\rho=0.319\%\) at \(A_{\rm p}=0.50\) and \(0.523\%\) at
\(A_{\rm p}=1.25\). The instantaneous-transverse and local-feedback contrasts
remain within \(0.024\%\) of coupled CS throughout.
These small contrasts are consistent with the nearly equal raw \(D_\rho\) and
\(D_v\) values above.

We distinguish the models more clearly using the geometric distances. With the same
endpoint averaging convention, their circulation and longitudinal values are
\begin{align}
 D_\omega={}&(3.65\times10^{-19},\,3.033\times10^{-4},\nonumber\\[-2pt]
              &\hspace{1.3em}3.042\times10^{-4},\,3.034\times10^{-4}),
              \nonumber\\
 D_\parallel={}&(3.030\times10^{-4},\,3.47\times10^{-19},\nonumber\\[-2pt]
                 &\hspace{1.3em}1.231\times10^{-6},\,3.566\times10^{-6}).
 \label{eq:raw_endpoint_geometry}
\end{align}
The reciprocal velocity is a gradient and therefore has numerically zero
\(D_\omega\), whereas the instantaneous-transverse velocity is divergence
free and has numerically zero \(D_\parallel\). Consequently,
\(D_\omega^{(\mathrm{rec})}\simeq0\) gives
\(\Delta_\omega^{(\mathrm{rec})}=-100\%\), indicating the absence of
circulation separation relative to the nonzero coupled-CS reference. In
Table~\ref{tab:pathseparation}, the instantaneous-transverse
\(\Delta_\omega\) ranges from \(-0.009\%\) to \(-0.017\%\), while the
local-feedback value ranges from \(0.132\%\) to \(0.355\%\) as
\(A_{\rm p}\) increases. These are relative contrasts with coupled CS, not raw
circulation distances. Local-feedback and coupled-CS dynamics are the two
models with nonzero endpoint traces in both geometric channels, although only
the latter evolves an independent connection according to the spatial CS
equation tested above.

In Table~\ref{tab:path}, we follow zero-input relaxation from the end of
the second pulse; the standard endpoint is at \(t_{\rm rel}=0.18\).
Within each seed, we average raw distances over widths and amplitudes
before forming contrasts and ratios. The large
reciprocal \(Q_\parallel\) reflects the different longitudinal scales of the
two models, not a large absolute coupled-CS denominator. At
\(t_{\rm rel}=0\), the raw pulse-grid means are
\(D_\parallel=3.13\times10^{-4}\) for reciprocal dynamics and
\(3.70\times10^{-6}\) for coupled CS, whose ratio is \(84.6\). 

In Table~\ref{tab:path}, we instead average the ten seedwise ratios and obtain
\(159.6\pm61.5\); the larger value and uncertainty arise because several
seeds have especially small coupled-CS denominators. During the
\(0.72\)-unit zero-input continuation, all raw separations decrease because
the relaxation term \(-\tau\rho\) remains active. The normalized quantities
need not decrease monotonically, since their coupled-CS reference values also
evolve; here we use them to compare the channel structure across models.
Reciprocal dynamics retains the pulse-order trace only in the longitudinal
channel, instantaneous-transverse dynamics only in the circulating channel,
and local-feedback and coupled-CS dynamics in both. The decay of the raw
separations therefore shows that pulse order remains distinguishable for a
finite time but is progressively forgotten, as expected for fading rather
than permanent memory.

\subsubsection{Linear decodability of pulse order}

Having established that the two histories remain distinct, we use the
classifier in Fig.~\ref{fig:readouts}(b) to test whether a simple linear
readout can recover their order. Density-only accuracy ranges from \(0.79\) to
\(0.83\), with overlapping uncertainties. Density therefore records pulse
order in all four models but does not isolate the CS mechanism.

The geometric feature families reveal the expected channel structure.
Circulation-only accuracy is at chance for reciprocal dynamics (\(0.51\)),
but reaches \(0.80\) for both instantaneous-transverse and coupled-CS
dynamics, which contain the rotated-gradient channel of
Eq.~\eqref{eq:leading}. Longitudinal features provide complementary
information and give an accuracy of \(0.78\) for coupled CS. Combining all feature families, we obtain accuracies of \(0.81\pm0.02\), \(0.80\pm0.02\),
\(0.87\pm0.02\), and \(0.86\pm0.02\) across the four-model ladder. In
paired seedwise comparisons, coupled CS exceeds the instantaneous-transverse
control by \(0.065\pm0.019\), whereas its difference from local feedback is
\(-0.012\pm0.019\). Access to both circulating and longitudinal traces thus
improves decoding relative to the instantaneous-transverse control, but we do not resolve an improvement relative to local feedback. At \(24^2\),
the corresponding paired differences are \(0.073\pm0.018\) and
\(-0.010\pm0.019\); fixed-input time-step refinement reproduces the
baseline scores. The comparison therefore persists under both refinements.

\begin{figure*}[!ht]
 \includegraphics[width=\textwidth]{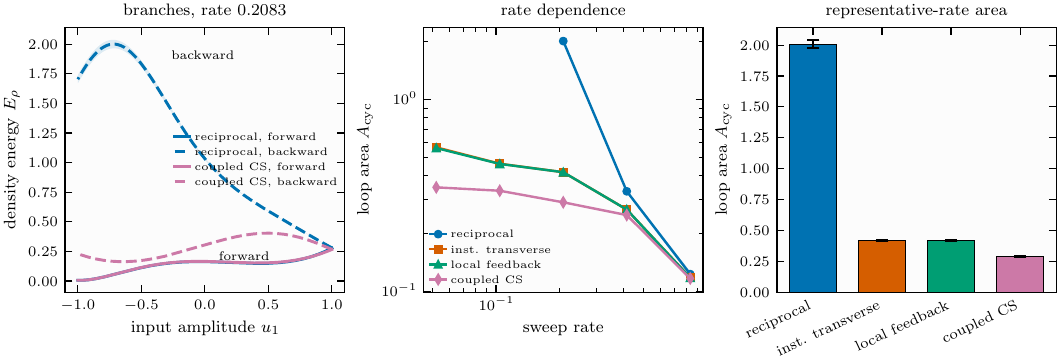}
 \caption{First-cycle density lag without pre-equilibration. (a) Mean forward (solid) and backward
 (dashed) branches at \(r_s=0.2083\); bands show SEM. We show only the reciprocal and
 coupled-CS branches because the instantaneous transverse controls
 overlap visually. (b) Loop area \(A_{\rm cyc}\) versus sweep rate. (c) Mean
 loop area with SEM error bars at the representative rate.}
 \label{fig:ladderhysteresis}
\end{figure*}

\subsection{Cyclic-drive density lag}

In the final reservoir protocol, we measure how closely the density
retraces its outward response when we reverse the input sweep. Figure
\ref{fig:ladderhysteresis}(a) shows the forward and return branches at the
representative rate, and panels (b) and (c) summarize their separation through
the loop area \(A_{\rm cyc}\) of Eq.~\eqref{eq:hystarea}.

At the fastest sweep, all four models have nearly the same loop area.
At the representative rate, the first-cycle means are \(0.292\) for
coupled CS, \(0.418\) for both instantaneous controls, and \(2.008\)
for the reciprocal model; the latter becomes unstable at the two slowest
rates. These unrelaxed first cycles include initialization effects.

When we repeat the representative cycle without resetting the states,
we find that the ordering reverses. On cycles two and three, coupled-CS mean areas are
\(0.271\) and \(0.266\), compared with \(0.209\) and \(0.214\) for
the instantaneous-transverse control (local feedback: \(0.209\) and
\(0.215\)). Grid refinement changes these means by less than \(0.7\%\),
and time-step refinement preserves them to the reported precision.
Thus the smaller initial CS loop is preparation-dependent, not a persistent
lag advantage. Three cycles do not establish a stationary or quasistatic
limit.

\begin{figure}[!ht]
 \centering
 \includegraphics[width=\columnwidth]{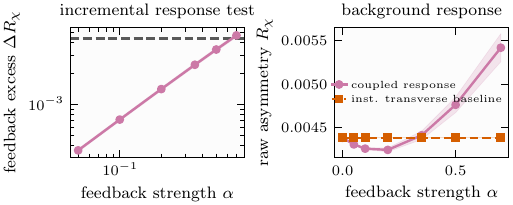}
 \caption{Incremental feedback response. (a) Background-subtracted excess
 \(\Delta R_\chi\); the dashed line gives the uncorrected \(\alpha=0\)
 scale. (b) Raw asymmetry \(R_\chi\) for coupled CS and the instantaneous
 baseline.}
 \label{fig:rate}
\end{figure}

\subsection{Feedback-dependent cross-response: \texorpdfstring{\(\Delta R_\chi\)}{Delta R-chi}}

In the final gauge-dynamics test, we isolate the contribution of advective
feedback to the two-site response. The raw response in Fig.~\ref{fig:rate}(b) is already
nonzero on the prepared inhomogeneous background. At \(\alpha=0\), coupled CS
gives \(R_\chi=4.39\times10^{-3}\), equal within rounding to the instantaneous
control, so this baseline cannot be attributed to advective feedback.

After subtracting the seedwise \(\alpha=0\) cross-sum, we find that \(\Delta R_\chi\) increases from \(3.62\times10^{-4}\) at
\(\alpha=0.05\) to \(4.72\times10^{-3}\) at \(\alpha=0.70\)
(Fig.~\ref{fig:rate}(a)). Fitting the four smallest nonzero strengths, we obtain a slope of \(0.983\pm0.002\), consistent with the first-order contribution of
\(\alpha\rho\bm v\) in Eq.~\eqref{eq:precurrent}. The incremental response is
small compared with the background at weak coupling and becomes comparable to
it at the largest \(\alpha\). This establishes a systematic feedback-induced
correction, while the raw coupled and instantaneous responses remain of the
same order.

\section{Conclusion}
\label{sec:conclusion}

We have addressed a mechanistic gap in field reservoir computing: how
input history is encoded in coupled density and geometric degrees of
freedom on an internal manifold. We formulate a conserved-current
evolution for a continuous density coupled to a CS connection. This
construction separates the connection component fixed by density through
Gauss law from an exact component generated by the history of nonlinear
transport. We thus give an explicit dynamical account of how the
transport field can retain information beyond its instantaneous
reconstruction from density.

We test this account through matched reciprocal, instantaneous-transverse,
local-feedback, and coupled-CS models. These comparisons distinguish the
roles of density-mediated transport, circulation, local nonlinearity, and
independent connection evolution. Neither the tested transverse
reconstruction nor the scalar local closure recovers the evolved exact
component. When we remove this component while preserving density and
Gauss-law curl, the subsequent density responses separate under identical
inputs. The separation is small but persists with finer spatial and
temporal resolution. Within the specified gauge and drift law, this
establishes that the additional connection state affects future transport.

We also find that dynamical relevance and computational benefit are
separate outcomes. All four models retain fading scalar memory and similar
density traces of pulse order, while their circulating and longitudinal
traces distinguish the underlying transport mechanisms. Coupled CS improves
pulse-order decoding over the instantaneous-transverse control, but remains
comparable to local feedback. Its smaller first-cycle lag also becomes
larger than that of both controls on subsequent cycles. These observations
show why neither a memory score nor an initial response comparison alone
identifies the contribution of an independently evolving field.

We establish these results at fixed domain topology with no harmonic
memory channel. They demonstrate gauge-organized dynamical memory without
establishing topological protection. The controlled removal of a connection
component does not establish naturally reachable histories with identical
density, and the closure comparisons do not exclude more general
density-based representations.

We propose the four-model comparison, field decomposition, controlled
component removal, and readout analysis as a strategy for examining a
broader class of coupled topological field and density models. Such
extensions require controls that preserve the relevant constraints while
separating instantaneous response from independent field evolution. The
central contribution is a link between the field equations and observable
memory: we identify a component that retains current history, demonstrate
its influence on subsequent density dynamics, and assess computational
performance against matched density-based controls.

Our future works will explore whether distinct input histories can produce the
same density with different subsequent responses. We would also examine
which tasks, input timescales, and readouts benefit from the additional
connection state. Controlled variation of domain topology and harmonic
modes would allow us to test for a computational contribution from
topology beyond the current-driven mechanism identified here.

\section*{Author declarations}
\subsection*{Conflict of interest}
We have no conflicts to disclose.

\subsection*{Ethics approval}
Our study is theoretical and numerical and involves no human participants
or animals. Ethics approval is not applicable.

\subsection*{Author contributions}
\textbf{Jyotiranjan Beuria:} Conceptualization; Methodology; Software;
Formal analysis; Investigation; Validation; Visualization;
Writing (original draft); Writing (review and editing).
\textbf{Venkatesh H. Chembrolu:} Conceptualization;
Writing (original draft); Writing (review and editing).

\subsection*{Use of generative AI and AI-assisted technologies}
We used ChatGPT and Grammarly for language editing, and ChatGPT for
conceptual clarification and assistance with numerical validation code.
We retain full responsibility for the manuscript and results.

\section*{Data availability}
The numerical data and code supporting this study are available from the
corresponding author upon reasonable request.

\bibliography{references}

\end{document}